\documentclass[11pt,draftcls,onecolumn]{IEEEtran}
\usepackage{latexsym,graphicx,pstricks,psfig,amsmath}
\usepackage{amsfonts,amssymb,mathrsfs,amsbsy}
\usepackage{epsfig,natbib}
\include{commands}
\usepackage{enumerate}
\usepackage{cases}
\usepackage{url}
\usepackage{multirow} 
\usepackage{array,colortbl}
\usepackage{color}
\usepackage{textpos}
\usepackage{algorithm}
\usepackage{algpseudocode}
\usepackage{times}
\usepackage{setspace}
\usepackage{pbox}
\usepackage{rotating}
\usepackage{acronym}
\usepackage{longtable}

\begin{document}


\title{IoT-Based Wireless Networking\\ for Seismic Applications}

\author{Hadi Jamali-Rad$^{\ast}$, and Xander Campman
\thanks{The authors are with Shell Global Solutions International B.V., 2288 GS Rijswijk, The Netherlands. e-mails:\{hadi.jamali-rad, xander.campman\}@shell.com. Corresponding author: Hadi Jamali-Rad, phone: (+31)704473185, e-mail: hadi.jamali-rad@shell.com. Part of this work has been filed as a patent in \cite{PatentWireless17}.}}

\markboth{Geophysical Prospecting}{Geophysical Prospecting}

\maketitle

%
%
%
%

\begin{abstract}

There is growing pressure from regulators on operators to adhere to increasingly stricter regulations related to the environment and safety. Hence operators are required to predict and contain risks related to hydrocarbon production and their infrastructure in order to maintain their license to operate. A deeper understanding of production optimization and production-related risk requires strengthened knowledge of reservoir behavior and overburden dynamics. To accomplish this, sufficient temporal and spatial resolution is required as well as an integration of various sources of measurements. At the same time, tremendous developments are taking place in sensors, networks, and data analysis technologies. Sensors and accompanying channels are getting smaller and cheaper and yet they offer high fidelity. New ecosystems of ubiquitous wireless communications including Internet of Things (IoT) nowadays allow anyone to affordably connect to the Internet at any time and anywhere. Recent advances in cloud storage and computing combined with data analytics allow fast and efficient solutions to handle considerable amounts of data. This paper is an effort to pave the way for exploiting these three fundamental advances to create IoT-based wireless networks of seismic sensors. 

To this aim, we propose to employ a recently developed IoT-based wireless technology, so called low-power wide-area networks (LPWANs), to exploit their long range, low power, and inherent compatibility to cloud storage and computing. We create a remotely-operated minimum-maintenance wireless solution for four major seismic applications of interest. By proposing appropriate network architecture and data coordination (aggregation and transmission) designs we show that neither the low data-rate nor the low duty-cycle of LPWANs impose fundamental issues in handling a considerable amount of data created by complex seismic scenarios as long as the application is delay-tolerant. In order to confirm this claim, we cast our ideas into a practical large-scale networking design for simultaneous seismic monitoring and interferometry and carry out an analysis on the data generation and transmission rates. Finally, we present some results from a small-scale field test in which we have employed our IoT-based wireless nodes for real-time seismic quality control (QC) over clouds.  
\end{abstract}

\section{Introduction and Related Work}
Wireless communications has been the driving force behind one of the largest and most successful sectors of industry during the past three decades. The telecommunications industry has become very mature from both technological and infrastructural perspectives; nowadays, it is possible to handle a gigantic amount of data transmission and coordination in stable and efficient ways. With the development of 3G and 4G technologies, and 5G on the way, wireless communications has proven to be capable of handling complex transmission media and mobility in a robust fashion. Obviously, there is a tremendous potential in wireless communications which is applicable to several other industries. 

The Oil and Gas industry as a whole and seismic applications in particular are good examples where an efficient data aggregation, transmission and storage are challenging due to the size of data and/or complications imposed by the environment. As an example, the technical and operational difficulties associated with scaling up cable-based land seismic operations motivate incorporating the potentials of wireless sensors in today's seismic activities. This is because cables are vulnerable to environmental effects and can create interference on neighboring cables. On the contrary, wireless sensors can be lighter and maybe cheaper per channel than the cable-based sensors but also dramatically easier to transport, install and retrieve. Besides, no cables involved decreases the excessive weight of cable-based seismic recording systems and makes transportation of the nodes cheaper, and at the same time rules out issues such as tangling and cable-break repairs. More detailed comparison between wired and wireless seismic systems can be found in (\cite{pellegrino2012wireless, Kendall15, daug2009evolution, Hollis05}). These advantages make wireless sensors a good choice for the following situations: first, locations difficult to reach and navigate, such as dunes, Jebels, and mountainous areas such as foothills and fold and thrust belts; second, temporary installations or surveys which are supposed to move rapidly to another location; third, remote locations with limited or no power access through wires. Therefore, immediate applications in our line of business such as quality control (QC) in harsh environments and monitoring in remote areas can benefit from networked wireless sensors. Notably, advanced wireless networking protocols and distributed data storage/processing can clearly add value and save us time and money. 

An overview of suitable wireless technologies in the market with the potential to be used in seismic operations as well as some networking ideas are provided in \cite{Gana08}. There are studies which particularly target ground motion and landslide monitoring or early warning systems for volcanic activities such as (\cite{Fleming09, Pereira14, Fischer09, Husker08, Weber07, Srinivas14, Picozzi10}); there are also those who focus mostly on exploration seismic acquisition (\cite{Tran07, Barakat08, Savazzi08, Savazzi09, Savazzi09_2, Savazzi09_3, Savazzi10, Savazzi11}). Each of these categories target different wireless technologies and networking designs based on what the nature of the scenario demands. Some ideas on using wireless sensor networks for seismic acquisition are sketched in \cite{Tran07} wherein different layers of the network design from different physical layer (PHY) technologies (WiFi, WiMAX, LTE. etc.) up to appropriate routing protocols are briefly discussed. Similar discussions with more information on network synchronization can be found in \cite{Barakat08}. 

The general idea in the literature for seismic data acquisition is to define clusters and sub-clusters of nodes where each node has limited range and capacity. There are also multiple higher capacity data-gathering (or relaying) nodes per sub-cluster in order to collect the data and then forward them through the other relaying nodes to a nearby gateway associated to each cluster. From the closest relaying node (so-called cluster-head) the data can be transmitted to gateways and then relayed through the gateways with a few hops until it reaches a central data storage/processing unit such as a truck nearby the acquisition site (\cite{Savazzi08, Savazzi09, Savazzi09_2, Savazzi11}). The idea inside each sub-cluster is to use short-range ultra-wideband (UWB) communication technologies employing multi-band OFDM (MB-OFDM) \cite{Batra04}. Such technologies offer high data-rates (around $100$ Mbps) in short ranges and an acceptable time-based self-localization possibilities in case global positioning system (GPS) is not available per node. It is proposed to use extended WiFi technology for the gateways to reach the storage/processing unit. Detailed discussions on the routing, medium-access-control (MAC) and self-localization of the network can be found in (\cite{Savazzi08, Savazzi09, Savazzi09_2, Savazzi09_3, Savazzi10, Savazzi11}). When it come to early warning systems for volcanic activities and ground motion monitoring, the number of sensors is typically substantially less compared to seismic acquisition because a limited number of sensors are spread over a large area. As a result, the general idea here is to employ long-range wireless technologies such as extended WiFi \cite{Weber07}, or some other long-range radio frequency (RF) technologies within the wireless local area networks (WLAN) family \cite{Husker08}. In the literature, these networks are typically laid out based on an extended star topology. At the center of each star there is a leading (or data-gathering) node, and these leading nodes are typically connected through high bit-rate digital subscriber line (HDSL) or fiber optic connections. 

The downside of the proposed approaches in the literature is that typically such high data-rate technologies are relatively expensive and also too power-consuming, especially when applied to large-scale networks. Their high data-rate in practice might not be necessary because most seismic applications can tolerate a reasonable amount of delay in transmission of data. While there is a large body of literature in the telecommunications sector on addressing all of the above challenges in great details within a different context, not much can be found on wireless network design for a broader range of seismic applications. What exists in literature is typically designed to fit a specific application or a location rather than a general study of the problem. Most of the existing works suggest using mature high data-rate wireless technologies in the market without appropriate consideration of their imposed cost and power consumption load for large-scale networking. 

An important phenomenon we have observed and incorporated in this paper is the advent of low-power wide-area (LPWA) wireless technologies in the market. These technologies are the response of wireless communications and cellular networks to the upsurge of attention that the ``Internet of Things'' (IoT) has recently received. LPWA networks (LPWANs) are best suited for applications that require a low data-rate but have to typically transmit over a long range in a battery-limited mode. It turns out that for a range of seismic applications (detailed in the next section) we can actually live with the provided data-rates by LPWANs. As a result, their low price per module, reasonably long range, and low power consumption makes them promising options for our networking designs. The novelty of this paper is four-fold. First, we put LPWANs at the core of our networking design allowing us to exploit the potentials of IoT-based wireless networks. Second, we develop network architectures that are a natural fit to a combined IoT - cloud storage/computing framework, which as a result can also benefit from a wide variety of cloud-based data analytics techniques. Third, we propose a cross-layer networking approach fitting the duty-cycled paradigm of LPWANs where we formulate how the operational delay-tolerance of the network in terms of data delivery, the required data-rate and data frame structure are inter-related. We end up with appropriate closed-form formulas enabling us to compute an estimate of the required data-rate for the target wireless technology. Finally, we put all these into practice by designing an IoT-based large-scale wireless network as well as by conducting a small-scale field test.       

The rest of the paper is organized as follow. We, first highlight four potential application scenarios. We then propose two different categories of network architectures based on the two classes of LPWANs, one capitalizing on the existing cellular infrastructure, and the other one revolving around a hybrid of private-cellular networks. With the PHY network architecture in place, we move on to a cross-layer design in order to efficiently handle transmission of complex seismic recordings. Due to limited space we omit higher-layer networking aspects such as network synchronization, localization, and data storage. We highlight that clouds are a natural choice for our data storage/analysis, given the fact that our target wireless technology LPWA is built to fit into a combined IoT-cloud computing/storage platform. Next, we look into how our networking designs could be applied to a sensor network for simultaneous seismic monitoring and interferometry over a producing field. We analyze the data generation rates of the network to show that they can be handled by the LPWANs. We also present some rough cost estimates for network setup and maintenance. Finally, we briefly present proof-of-concept field-test results for seismic QC with wireless nodes.  

\section{Scenarios of Interest: Seismic Setup and Data Generation}
\label{sec:scenarios}
There are four major scenarios of interest in this paper, which are of practical importance in the seismic domain. The corresponding systems might generate intermittent (triggered) seismic data as in earthquake monitoring systems, or sometimes their data should be recorded continuously to create meaningful maps, as in seismic interferometry. In the following, we briefly explain our scenarios of interest, and roughly outline their areal coverage and data generation size. Notably, we believe that there are several other applications in the Oil and Gas industry where our wireless networking ideas can be applied with minimum modification. 

\subsection{Ground Motion Monitoring (GMM)}
\label{ssec:motion}
%
\begin{figure}[!t]
\centering
\includegraphics[width=0.8\textwidth]{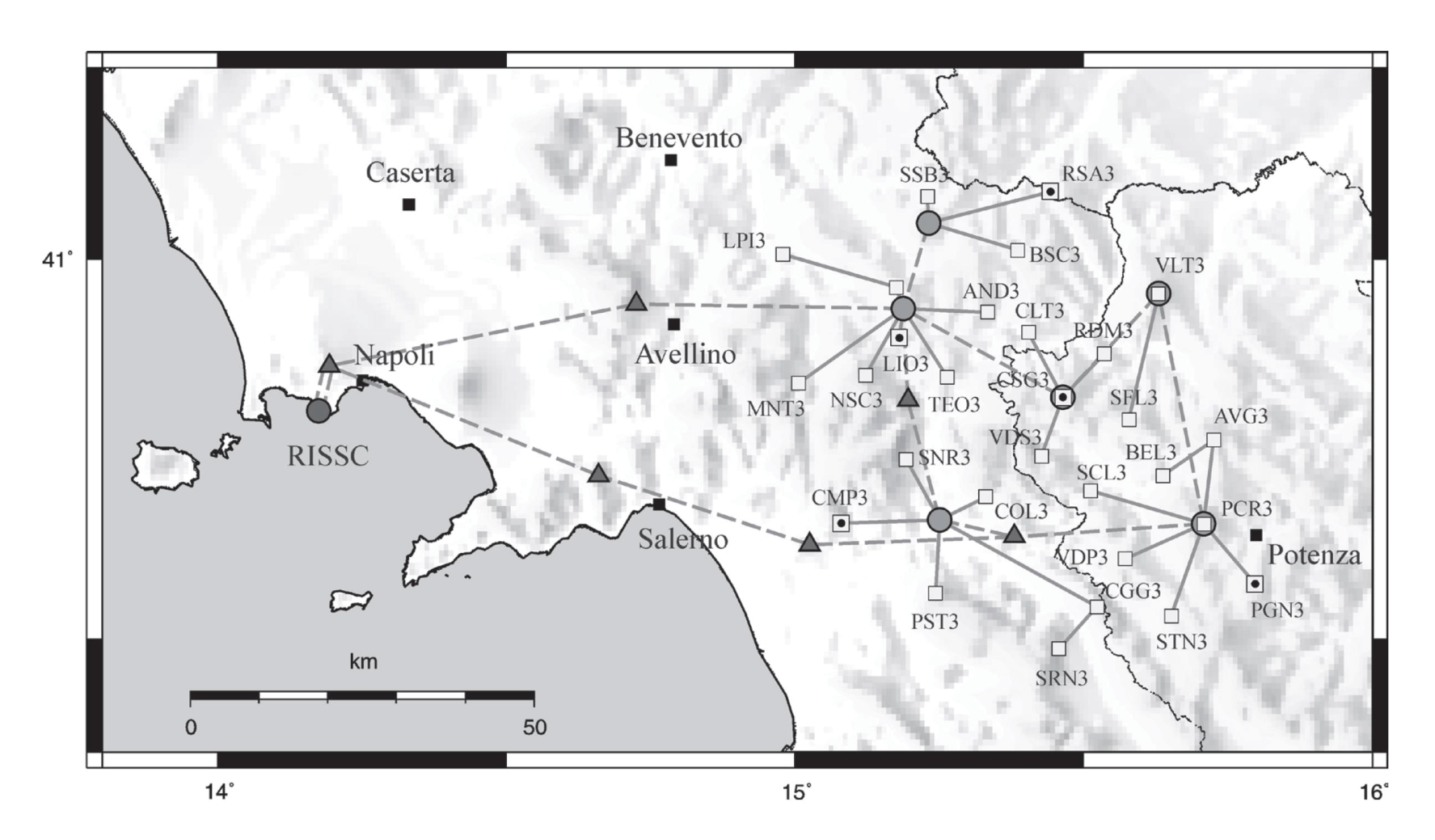}
\caption{Ground motion monitoring network in the southern Apennines \cite{Weber07}. The squares represent the network stations and the circles the local control centers (LCCs). The gray lines are the radio links between the stations and the LCCs, and the dashed lines are planned synchronous digital hierarchy (SDH) carrier-class radio upgrades for early-warning applications. The triangles represent radio repeater points.}
\label{fig:grndMot}
\end{figure}
Even though ground motion monitoring (GMM) is not traditionally an application in the area of exploration seismology, it is still a relevant seismic application for which considerable amount of wireless studies have been conducted. Therefore, we can learn valuable lessons from such studies (\cite{Fleming09, Pereira14, Fischer09, Husker08, Weber07, Srinivas14, Picozzi10}). GMM is an on-demand operation in which sensor nodes (geophones or accelerometers) continuously observe signals but only start recording data when an event (surpassing pre-defined threshold magnitude) occurs. This leads to a trigger based on which the nodes start recording data and this data needs to be kept for later analysis or be transmitted for immediate actions in early-warning systems. Sometimes the nodes keep on recording and overwriting their readings for a period of time in their memory buffer and once they are triggered the data corresponding to a short time-span before and after the trigger are kept as useful data and/or transmitted. The size of data to be transmitted per node is a product of the number of triggers by the size of recorded data per trigger. The recorded data should have a good resolution (high sampling rate and also high number of bits per sample) in order to ensure accurate source mechanism and source location estimation. GMM scenarios are not so demanding in terms of data generation/transmission because triggers do not happen very often. With regards to setup, such scenarios are typically comprised of a (few) tens-to-hundreds of sensors spread over an area of a few (hundreds of) kilometers squared. This means on average spacing of the adjacent nodes can be of the order of kilometers which is a challenging parameter as far as wireless transmissions are concerned. Fig.~\ref{fig:grndMot} depicts a wireless network design in the Southern Apennines (Italy) for an early earthquake warning system. The wireless network enables us to efficiently transmit the triggered sensor data and thus make fast and appropriate decisions for alarming systems. We should highlight that here the scenario of interest for us is still delay-tolerant and does not necessarily have to transmit all the data very fast and immediately.  

\subsection{Ambient Noise Seismic Interferometry (ANSI)}
\label{ssec:interfer}
%
\begin{figure}[!t]
\centering
\includegraphics[width=0.8\textwidth]{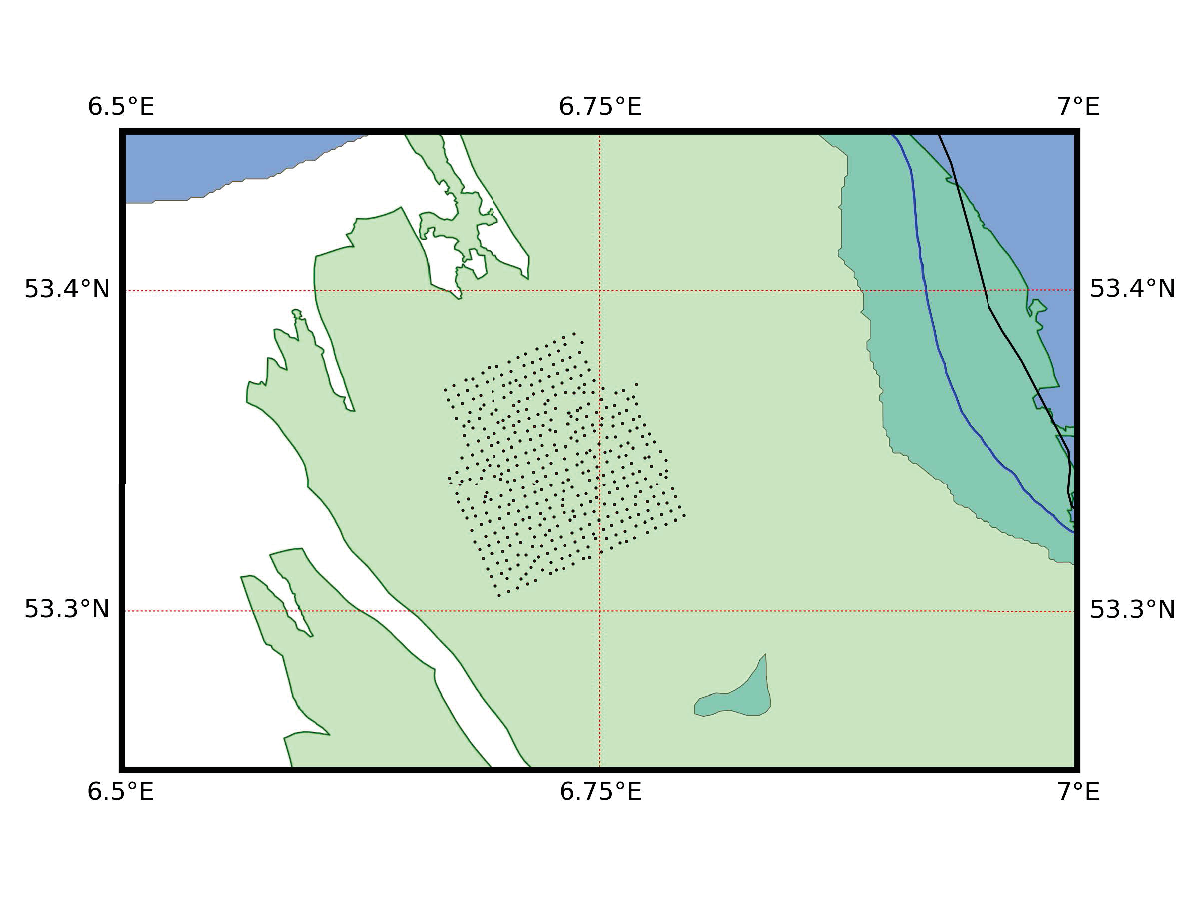}
\vspace{-0.2cm}
\caption{Roving seismic interferometry network operated by NAM in Groningen, the Netherlands. The dots represent $3$-component sensors, and the green area on the map is the outline of the Groningen field.}
\label{fig:interf}
\end{figure}
Ambient noise seismic interferometry (ANSI) allows geophysicists to gain important information about shallow subsurface, as well as to estimate and remove ground roll. ANSI can be applied to both passive and active seismic scenarios. It utilizes the cross-correlations of signals at different receiver pairs to reconstruct the Green's functions of the subsurface between the receiver pairs. The theory of seismic interferometry is based on the seminal work presented in (\cite{Campillo03, Shapiro04}). In practice, these cross-correlations should be computed over a long time (large span of continuous-time recordings) in order to converge to Green's function. It means, in contrast to GMM and early warning systems, seismic interferometry requires the sensors to keep on recording their noise readings. These recordings should then be transmitted to a fusion center (FC) for creating maps and for the analysis of the subsurface. It is preferred that this continuously recorded data can be transmitted in a real-time fashion or regularly for subsurface monitoring purposes. This highlights the importance of an appropriate wireless network design to efficiently handle the delivery of the data. The noise can be recorded with a low sampling rate as well as a low resolution (number of bits per sample) and still provide the required information from the shallow subsurface when cross-correlated. This fortunately keeps the data generation at a reasonably low rate. The challenge here is mostly due to the continuous recording and transmission of data. Regarding the setup, interferometry networks are typically dense networks containing thousand(s) of nodes deployed over a region of many squared kilometers. As a result, spacing among neighboring nodes can be as large as few hundred meters up to about a kilometer. Fig.~\ref{fig:interf} depicts a roving interferometry network consisting of more than $400$ $3$-component sensors in Groningen, the Netherlands. 

\subsection{Microseismic Fracture Monitoring (MFM)}
\label{ssec:frac}
%
\begin{figure}[!t]
\centering
\includegraphics[width=0.5\textwidth]{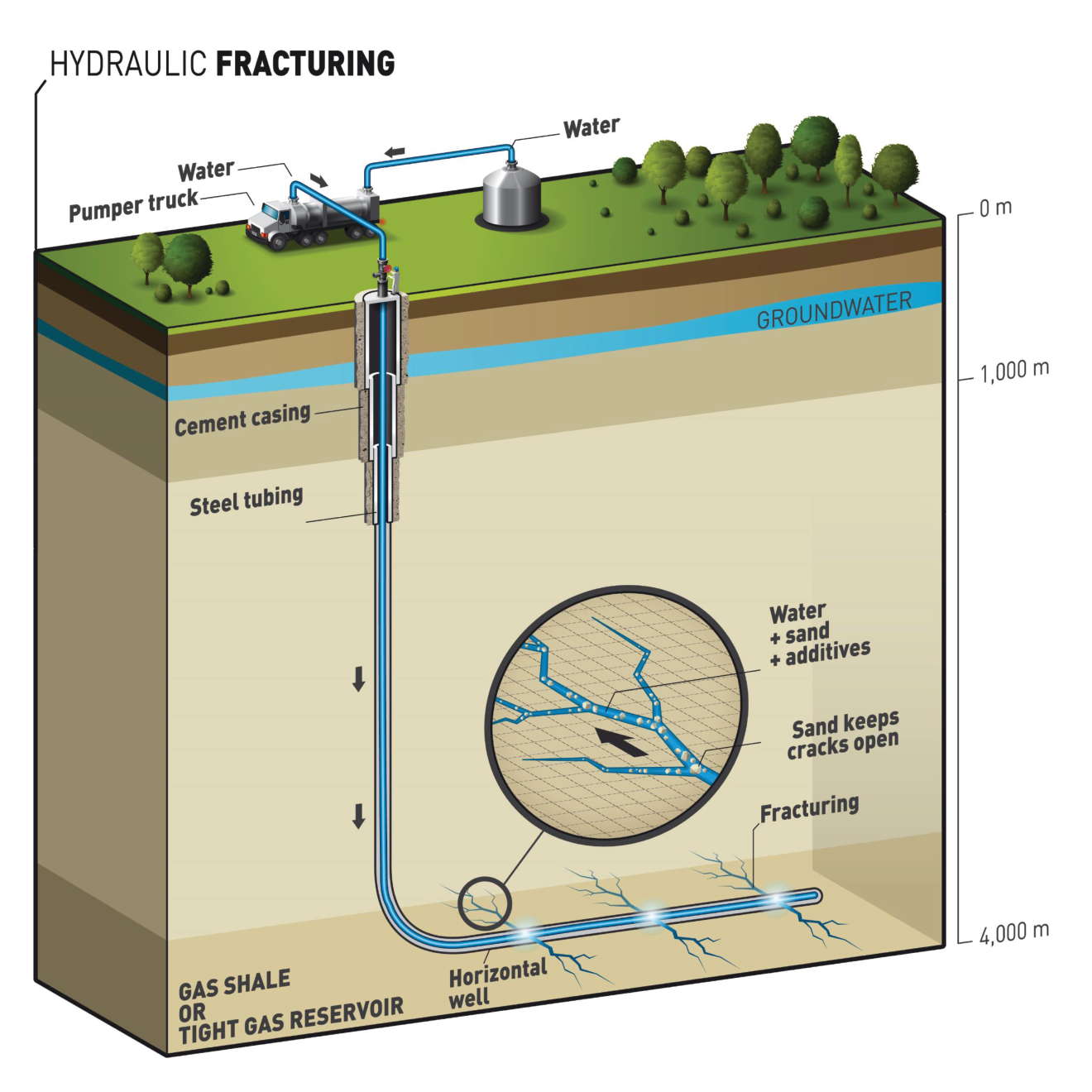}
\caption{A typical hydraulic fracturing setup where a high-pressure fluid is injected into a wellbore to create cracks in deep-rock formations through which oil and/or gas will flow more freely \cite{HydroFracImage}.} 
\label{fig:fracMon}
\end{figure}
Microseismic fracture monitoring (MFM) provides important information about volumetric stress/strain and failure mechanisms in reservoirs and thus helps to analyze and affect the productivity level of wells using hydraulic fracturing (\cite{GP14, LeCampion04}). The hydraulic fracturing process is shown in Fig.~\ref{fig:fracMon}. Sensors used for microseismic monitoring can be installed in a borehole as well as at the surface. Here, we focus on sensor networks deployed on the surface for microseismic monitoring. The area of interest for fracture monitoring typically extends to about a kilometer squared and sensors are normally placed every $50$ to $100$ meters depending how accurate moment tensors and location information of fractures should be estimated. What is recorded per sensor is the full trace data in a continuous fashion typically for a few hours. In some scenarios sensors are only triggered when the hydraulic material is injected in order to create fractures. In such a case recorded data should be transmitted continuously or in a real-time fashion to be able to keep track of fracture extension. From data size (per sensor) perspective this scenario is relatively demanding. However, the total amount of data is not huge as the number of sensors is relatively low compared to the other scenarios of interest.   

\subsection{Quality Control for Active Land Seismic Surveys (QCLS)}
\label{ssec:qc}
%
\begin{figure}[!t]
\centering
\includegraphics[width=0.75\textwidth]{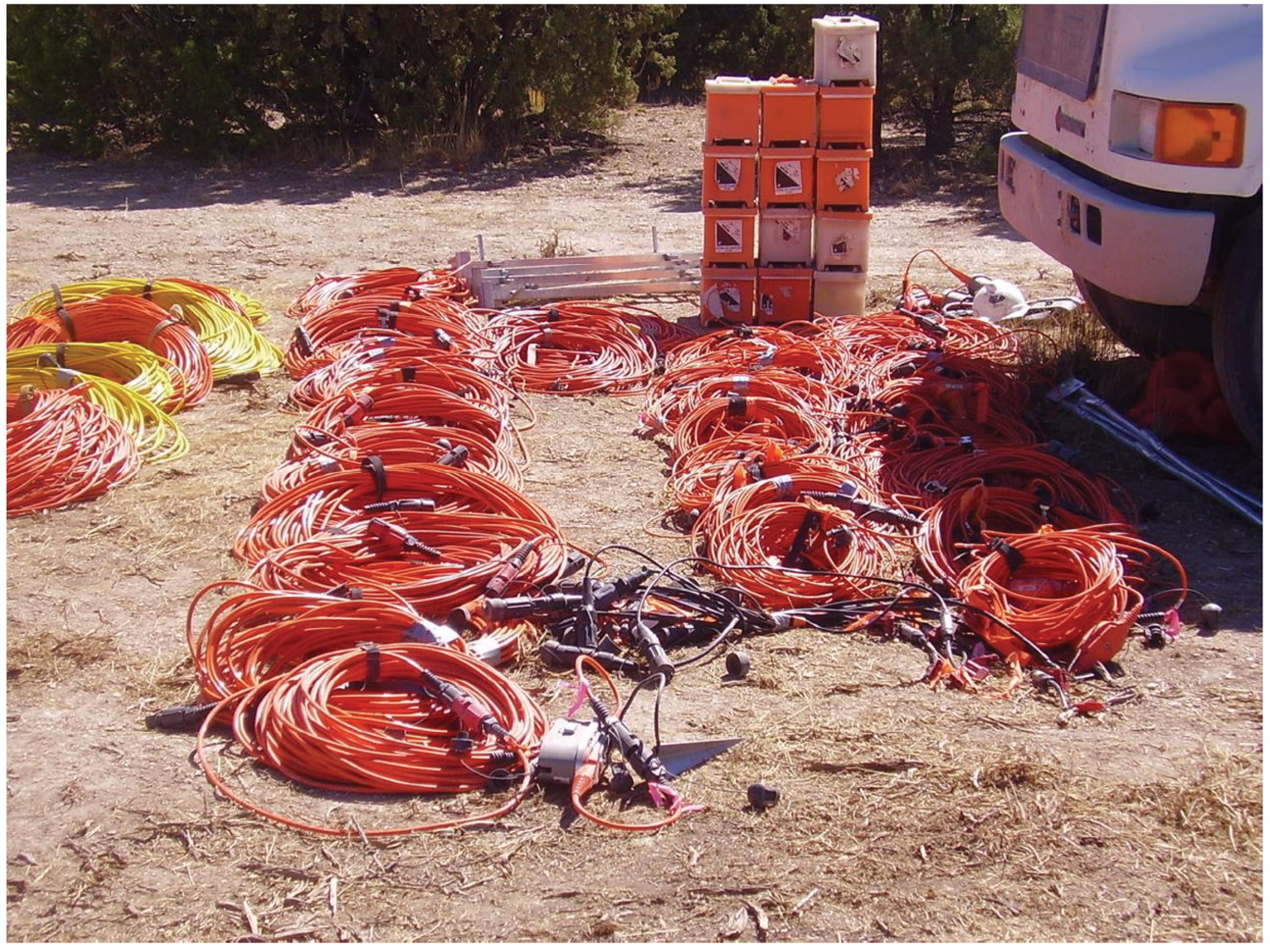}
\vspace{-0.4cm}
\caption{Traditional cable-based seismic acquisition with tens of thousands of nodes connected through cables \cite{Kendall15}.}
\label{fig:SeisACQ}
\end{figure}
Active seismic surveys are large and lengthy undertakings. Areal coverage of an active spread is roughly $100~\text{km}^2$. Spacings of sensors along the receiver lines (in-line direction) is typically $12.5$ m to $25$ m and the spacing of receiver lines in the other (cross-line) direction is about $200$ meters. This means there can be tens of thousands to even hundred thousands of sensors in place. Fig.~\ref{fig:SeisACQ} shows a traditional cable-based seismic acquisition scenario. Acquisition crew control the quality of the acquisition through a few parameters for every shot on a selection of station every once in a while, typically on a daily basis. These parameters include root-mean-squared (RMS) noise level per channel, geophone tilt, etc. It is of high interest also if sensors can report QC alerts, such as theft (based on significant change of their locations), or unexpected battery issues. In contrast to the MFM, here the size of data per sensor is only a few bytes in the worst case whereas the number of sensors is huge. This besides relatively close spacing of sensors along the in-line direction necessitates a different approach for collecting, coordinating and transferring the data through a wireless network. Sometimes QC is conducted on a per-shot-basis wherein only a subset of nodes could be monitored for each shot. If properly incorporated in data communication protocol, this inherent per-shot scheduling helps to reduce the communication burden. 

Table~\ref{tab:scenSumTab} summarizes the main features of the four scenarios of interest we discussed in this section. In an on-demand operation, the data is recorded intermittently based on a trigger whereas in a continuous operation data is recorded continuously.    

\begin{table}[!b]
\renewcommand{\arraystretch}{1.2}
\caption{Scenarios of Interest and Their Features}
\label{tab:scenSumTab}
\centering
{\footnotesize
\begin{tabular}{| m{1.5cm} | p{3.0cm} | p{2.5cm} | p{2.5cm} | p{3cm} |}
\hline
\textbf{Scenario} & No. of Nodes &  Spacing & Area Covered & Type of Operation\\
\hline
\hline 
GMM & $10-100$ & $1-20$ km & $100 - 1000\, \text{km}^2$ & On-demand \\
\hline 
ANSI & $1000-10,000$ & $100-1000$ m & $1-100 \, \text{km}^2$ & Continuous \\
\hline
MFM & $100-500$ & $50-100$ m & $1 \, \text{km}^2$ & On-demand/continuous \\
\hline
QCLS & $100,000 - 500,000$ & $10-200$ m & $100 \, \text{km}^2$ & On-demand/continuous \\
\hline
\end{tabular}}
\end{table}  
%

\section{Suitable Wireless Technologies in the Market}
\label{sec:wirTechs}

Selecting an appropriate wireless technology for the PHY layer of the wireless data communication network is an important design concern because it involves several factors to be taken into account. For instance, the size of the covered area and the density of the sensors define the required communication range of the wireless technology. From this angle, wireless technologies are divided into two categories, short-range (roughly up to a few hundred meters) and long-range (from a kilometer up to a few tens of kilometers). The category of choice really depends on our design criteria, infrastructure in the area of interest, and environmental parameters \cite{Gana08}. 

Another important parameter is the frequency band permissions, as well as required data-rate. The former is important because some technologies are allowed to work in specific unlicensed frequency bands (UB) which are free such as the industrial, scientific and medical (ISM) band. Some technologies operate in licensed bands (LB), and thus, they need subscription. The technologies in the UB have the advantage of being free in terms of frequency usage; however, they have to compete with the other active devices operating on the same frequency band in the vicinity and might experience interference issues that affect their performance. Technologies operating in both LB and UB are fine in our case (LB is preferred because of being congestion-free); they just have to provide a middle range of data-rates because our scenarios of interest are not highly data-demanding, as we discussed in Section~\ref{sec:scenarios}.

Finally, power consumption and price per module are important characteristics when it comes to setting up a full network. The wireless technology should be selected so that the whole network satisfies the total available cost budget, as well as power consumption budgets (per module and total). Power consumption budget becomes critical especially when there is no power source available at the sensors, and thus sensors should rely on their own battery or other means of power such as solar panels. From this perspective, traditional cellular machine to machine (M2M) communication standards (based on for instance 3G/4G) are rather expensive and they consume too much power for large-scale minimum-maintenance operations. Given the fact that short-range technologies (such as WiFi, ZigBee) do not seem to be a feasible option for us too, we then look for technologies that are power-efficient and cheaper than cellular M2M but also long-range to suit our purpose. The answer to this quest has recently become available in the market known as the family of LPWA technologies. 

\subsection{Low-Power Wide-Area Wireless (LPWA) Technologies}
\label{ssec:lpwa}
%
\begin{table}[!t]
\centering
\includegraphics[width=0.8\textwidth]{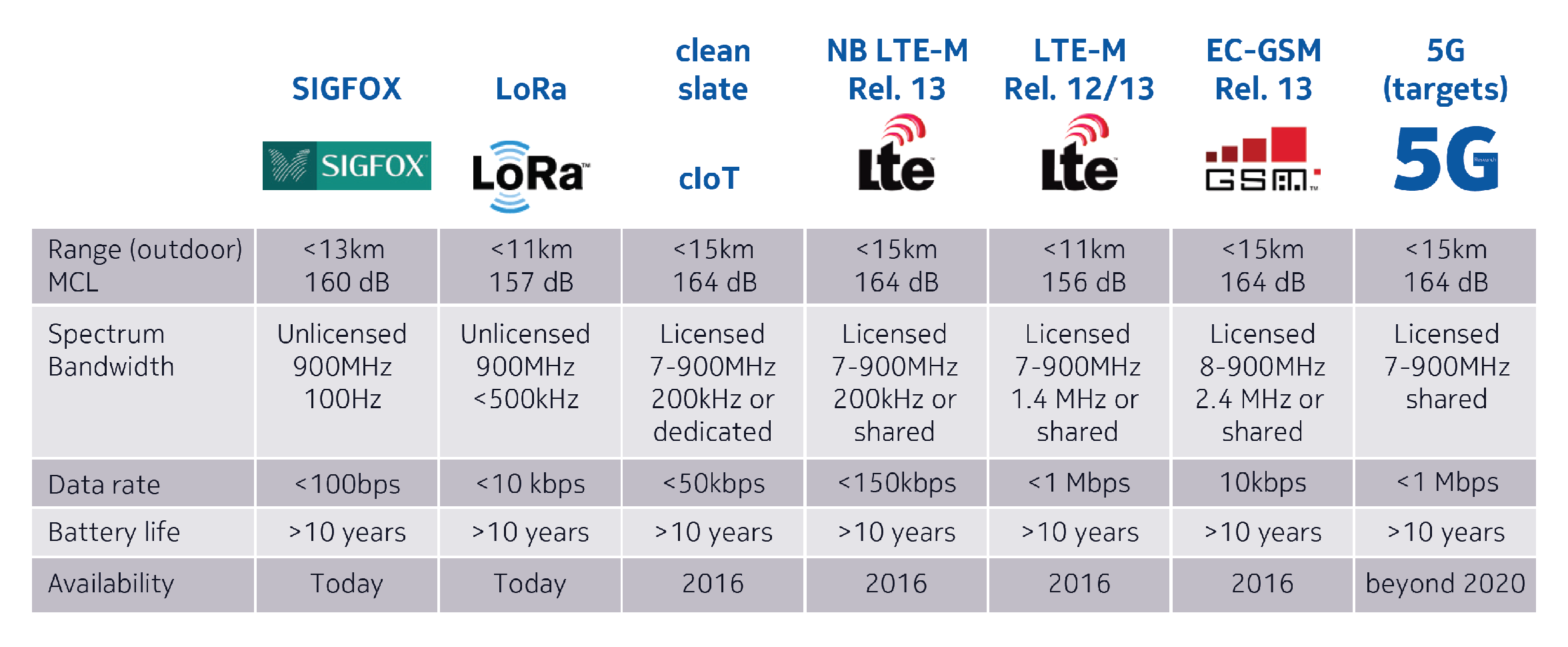}
\vspace{-0.3cm}
\caption[LPWA technology overview]{LPWA technology overview \cite{NokiaWP}.}
\label{tab:LPWAs}
\end{table}
LPWA's can be categorized into two separate categories. On one hand, there are available LPWA technologies such as Sigfox and long-range (LoRa) which operate in the UB. On the other hand, for the sake of a relatively larger data-rate, there is growing interest in modifying cellular standards like third generation partnership project (3GPP) and long term evolution (LTE) specifically for IoT applications. This second category is sometimes referred to as cellular IoT (\cite{NokiaWP}), and the technologies involved mostly operate in the LB. A key driver behind the growth of the second category is the existing infrastructure for cellular networks and that they could be used by cellular IoT's. These two categories of LPWA's are summarized in Table~\ref{tab:LPWAs}. In short, two classes of LPWAN's (i.e., cellular-based such as NB-LTE-M, EC-GSM, LTE-M and private ones such as Sigfox, LoRa) claim to offer a considerably long range of up to $15$ km in line-of-sight (LoS) conditions, extremely long battery lifetime (about $10$ years if the standard protocols are followed), and low but still tolerable data-rates ($10$ to $100$ kilo bits per second (kbps)) for our scope of applications. Note that NB and EC stand for narrow-band and extend-coverage, respectively. More importantly, note that NB-LTE-M is sometimes also referred to as NB-IoT; we use the latter naming convention in the following. 
\begin{figure}[!t]
\centering
\includegraphics[width=0.6\textwidth]{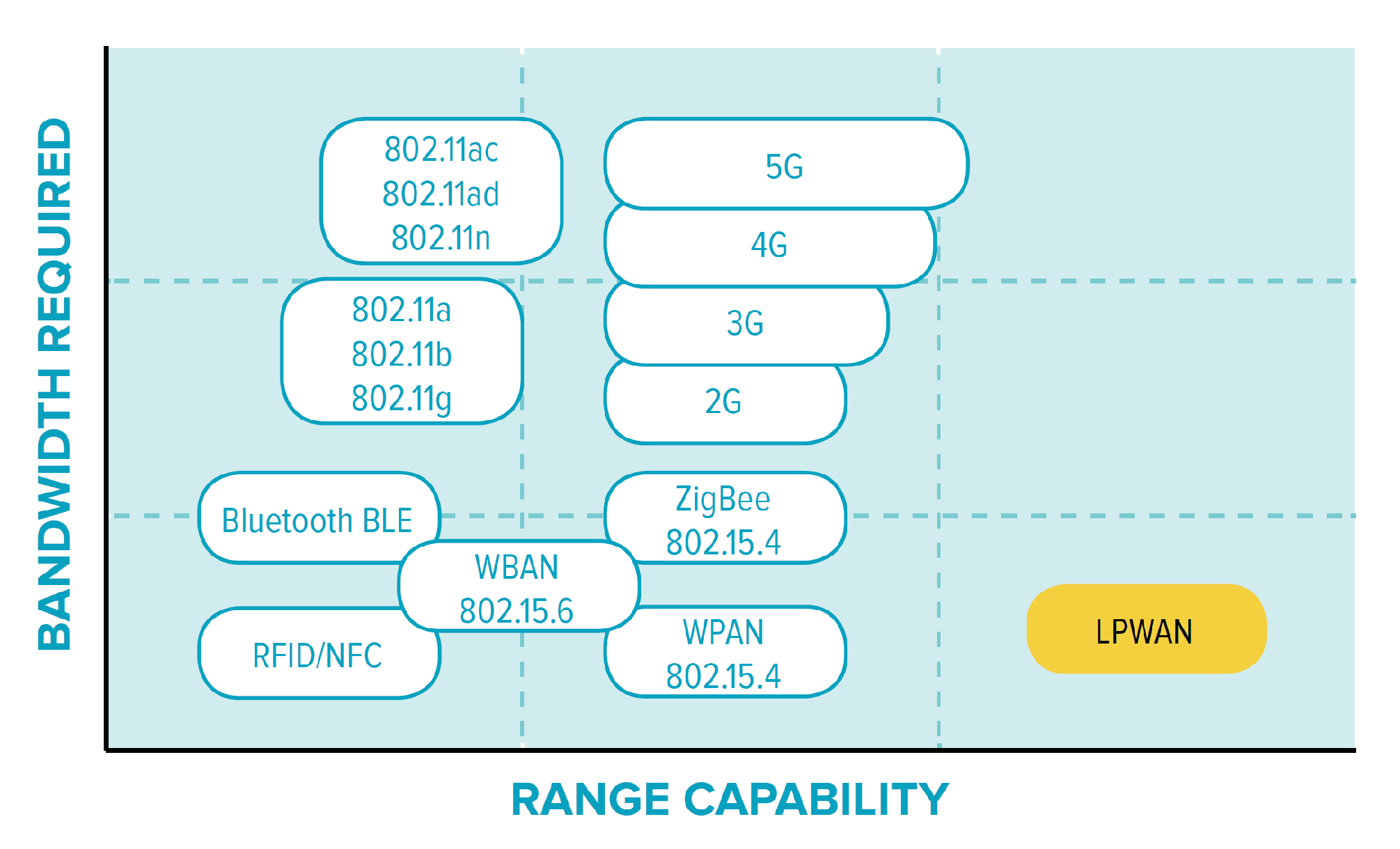}
\caption{LPWANs among the other legacy wireless technologies \cite{LinkLabsLPWAN}. LPWANs stand out as they offer a long rage but a relatively low data-rate as compared to the existing wireless technologies. }
\label{fig:LPWANs}
\end{figure}

The corresponding networking protocol for LPWA's  is commonly referred to as LPWAN. The preferred network topology for LPWANs is the star or extended star topology. There are two major areas where LPWANs are best suited. First, for fixed nodes with medium-to-large densities such as smart lighting controllers, smart grids, etc.; second, long-life battery-powered applications such as smart agriculture, battery-powered access control points. This basically defines the long-range low-bandwidth applications as the sweet spot for LPWANs to fill in the gap of the existing technologies as shown in Fig.~\ref{fig:LPWANs}. Note that here 5G refers to the initial high data-rate standard whereas in Table~\ref{tab:LPWAs} it refers to a new amendment of 5G focusing on IoT applications. LPWANs achieve this higher range primarily by higher receiver sensitivities of around $-130$ dB compared to $-90$ dB to $-110$ dB in traditional wireless technologies \cite{LinkLabsLPWAN}. On the other hand, two general shortcomings of the LPWAN family are the followings: first, the ones operating in UB (such as LoRa) have to follow a certain limitation in terms of duty-cycle or maximum time-on-air depending on regional regulations, second, as a result of the narrow-band signal in this family they cannot offer a good time-based node localization accuracy \cite{LinkLabsLPWAN}. This means that for applications that require a good node localization capability using GPS is highly recommended. 

In the following, we show that if the target seismic application can tolerate delay in data transmission, the low data-rate and limited duty-cycle/time-on-air are not an issue. On the other hand, the combination of long range and low power consumption makes LPWANs a technology worthy of being considered for seismic operations. Our primary focus among the LPWAN family is on LoRa and NB-IoT technologies. 

\section{Physical Layer (PHY) Design}
\label{sec:phy}
As discussed earlier, among our scenarios of interest, wireless networking for seismic acquisition and wireless early warning systems received quite some attention in the literature. Bear in mind that we are mainly interested in QC for seismic acquisitions whereas handling the data itself is the ultimate goal in the literature. These two have different requirements in terms of delay tolerance and data volumes, but similar network size and sensor spacing requirements. This means there will certainly be shared features in their corresponding network designs. Our goal in this paper is to design PHY architectures that are affordable for large-scale implementation, they fit into our vision of combined IoT-cloud computing framework, and they can handle long-range transmissions for ``not-so-data-demanding'' scenarios of interest. When it comes to networking design for applications of our interest, the following remarks are in place:

\begin{itemize}
\item[-] Our proposed architectures try to be infrastructure-aware in the sense that whenever there is a possibility to tap into existing (H)DSL lines at public or state-owned buildings, we take the opportunity. Note that the downside is that such nodes cannot participate in the end-node wireless inter-communications directly but still could be coordinated through the network server.  
\item[-] For monitoring purposes, each node in the network should be able to record its observations in a buffer for at least a period of time $t_b$. The data generation rate and $t_b$ determine how much memory for the buffer is required. 
\item[-] Our schematic view of the network simplifies the concept of core and access networks \cite{TanenbaumCompNet} by considering a service center where fiber and DSL data come together and from there on get forwarded to the network server which can be on a cloud. This also fits into our general idea of establishing an IoT-cloud paradigm for sensing, storage and computation. Interested reader is referred to \cite{EricssonBackhaul} for more details on the backhaul and backbone network configurations. 
\item[-] From the network server, the data will be reachable at different application servers at different locations for monitoring, analysis, and decision making. 
\end{itemize}

In the following we propose PHY architecture designs using the two categories of LPWAN family; namely cellular network-based and private-network based architectures. 

\subsection{Cellular Network-Based Data Transmission}
\label{ssec:cellTrans}
%
\begin{figure}[!t]
\centering
\includegraphics[width=0.9\textwidth]{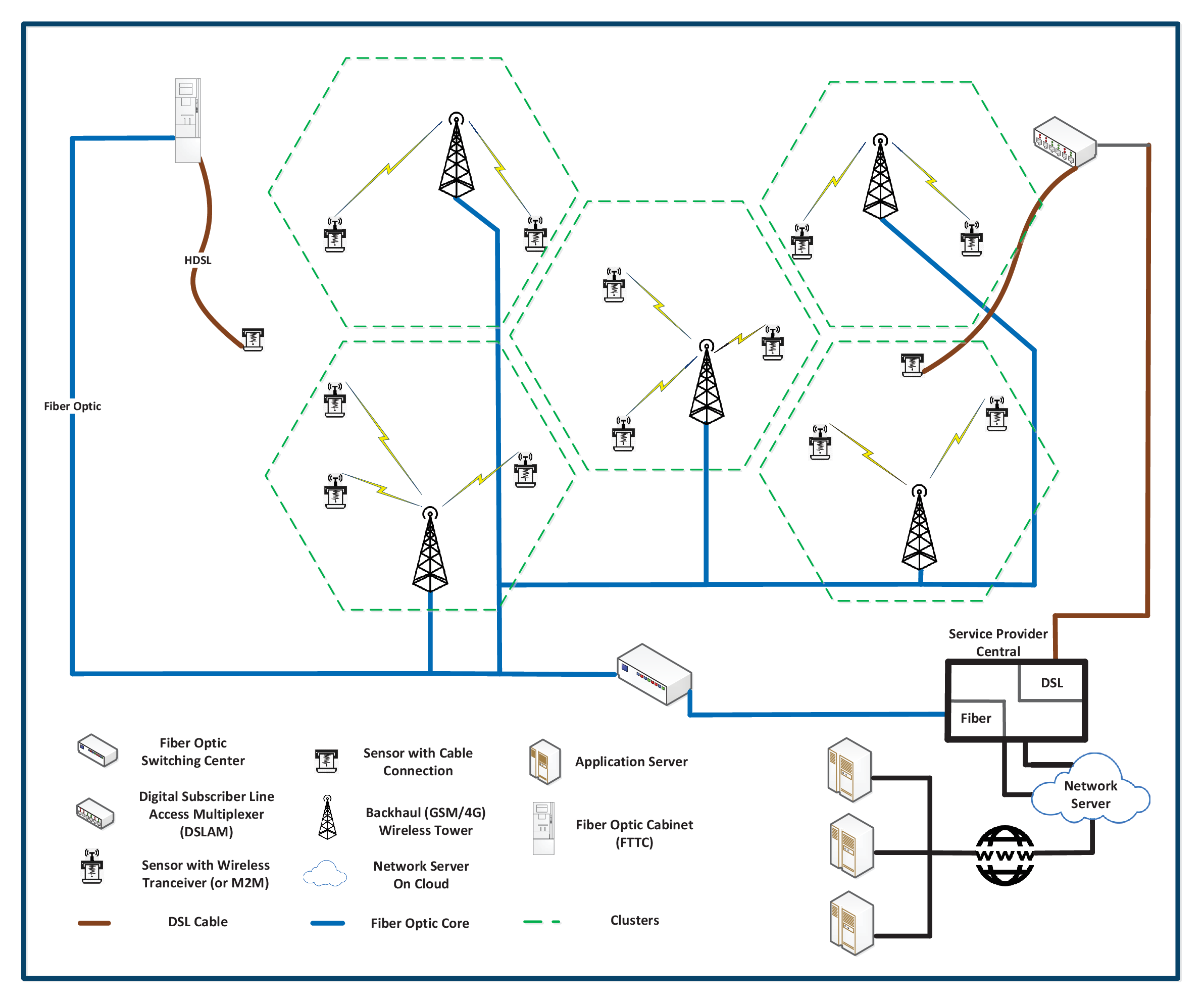}
\caption{Architecture I: cellular network-based data transmission focusing primarily on exploiting the cellular-network infrastructure in urban areas.}
\label{fig:Plan1}
\end{figure}
%
The network architecture depicted in Fig.~\ref{fig:Plan1} (Architecture I) is a large-scale network focused on primarily exploiting the cellular-network infrastructure. This means such a network is preferred to be deployed in areas where a established cellular-network infrastructure exists. Nowadays, most of urban and even suburban areas in Europe and USA have a rich cellular infrastructure. If the area of interest is not so far from existing cellular infrastructure, expanding the coverage of the network by creating edge cells and tapping into the capacity of the wireless backhaul is always an option. This, however, does not necessarily hold for regions lacking a proper cellular infrastructure, which is a limiting factor for this network architecture. The considerable capacity of the towers and in principle wireless backhaul makes this architecture flexible enough to scale up to thousands of nodes. Besides, the coverage area of each cell is typically of the order of tens of miles, which allows us to cover an immensely large area of interest.   

Each wireless-equipped node in Architecture I immediately communicates with a wireless tower or with an intermediate repeater base station (or a so-called Node-B mast) to reach a wireless tower in the backhaul network. From there on the data is forwarded through appropriate links (fiber, copper, or microwave) to a main gateway and then it will be available on the internet through the network server. At the application server side the user has the possibility to communicate back to each node either separately or to a group of them for instance based on the cell area in which they are located. Basically, in this architecture the cells around each wireless tower define our ``virtual'' clusters. Depending on the wireless technology employed at the end-nodes, the nodes can or cannot inter-communicate. If properly programmed, gateways can act as intermediate points in the network for coordination or decision making, or all the decisions can be made at the server side. Any cellular-based wireless technology that has a compatible transceiver module in order to tap into GSM/3G/4G wireless backhaul can be used in this scenario. This can be any M2M cellular standard or, preferably for our scenarios of interest, the technologies lying in the second category of LPWAN family, namely NB-IoT, EC-GSM, and LTE-M. Most of these technologies and their medium-access-control (MAC) standards provide the possibility to transmit a reasonable amount of data on a daily basis. Architecture I can handle a wide variety of scenarios including GMM, ANSI, and MFM, except those executed in isolated areas where the cellular network coverage does not exist. As an example, QCLS is typically executed in the areas lacking a established network backbone, and thus does not seem to be a good fit. 

\subsection{Hybrid Private-Cellular Network-Based Data Transmission}
\label{ssec:hybTrans}
%
\begin{sidewaysfigure}
\begin{center}
\includegraphics[width=0.99\textwidth]{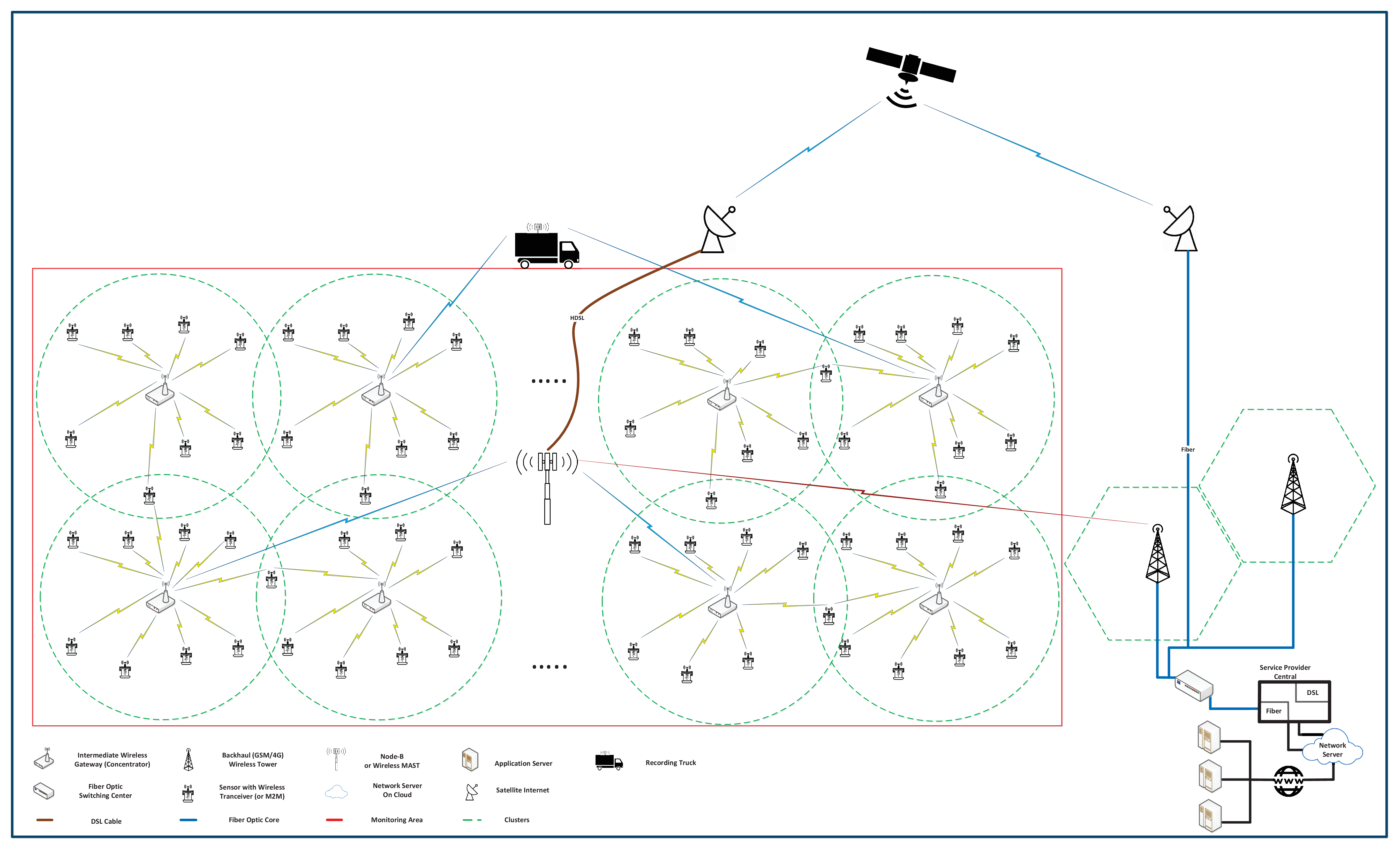}
\end{center}
\caption{Architecture II: hybrid private-cellular network-based data transmission focusing on a private networking especially in remote areas.}
\label{fig:Plan3}
\end{sidewaysfigure}

The network architecture depicted in Fig.~\ref{fig:Plan3} (Architecture II) is more focused on a private networking strategy especially in remote areas where a established cellular backbone is lacking. Here, groups of nodes create clusters around so-called data-gathering nodes with higher capacities as opposed to cellular network-based architectures (Architecture I) where each node tries to immediately transfer the data to wireless towers/gateways. This architecture is essentially devised to handle large number of nodes spread over large areas. The use of repeaters/concentrators imposes a limit on the capacity that can be handled per cluster, but also provides flexibility to make/apply decisions per cluster. Depending on the technology being employed the size and area of the clusters can vary. The network is supposed to be deployed far from a established cellular network infrastructure which mandates different ways of data aggregation and coordination. 
 
Here, the end nodes first communicate with their cluster-heads (so-called data-gathering nodes or concentrators). We consider three different options for the uplink (path from the end nodes towards gateways) communications from cluster-heads: 1) recording truck appropriately located right next to or in the middle of the area in order to collect all the data. The downside is that if the truck is not equipped with satellite communication interface, the data cannot be available on the cloud in a real-time fashion, 2) establish a satellite communication link at the isolated area in order to transfer the data to satellite receiver hubs and from there on (with fiber for instance) to the network server side, 3) create one or multiple cells by putting up extra intermediate wireless masts or Node-B's to tap into the wireless cellular infrastructure nearby (only if available). The rest of the path is the same as in Subsection~\ref{ssec:cellTrans}. The downlink path is also similar with the notable flexibility that part of the decision-making process can happen at the cluster level. To be more specific, if properly programmed, the cluster-heads can help to create a regional consensus among the cluster members before activating their transmitters and forwarding their actual data. This can for instance lead to a ``consensus triggering'' procedure for power-efficient node activation. The proposed architecture can be employed for different applications and the wireless technology should be accordingly adopted. Our main target here is the first category of LPWAN (such as LoRa) where bi-directional communication links can be established between nodes and concentrators and from the concentrators to any compatible wireless gateway. This architecture is flexible enough to handle on-demand and regular transmission scenarios. Architecture II is an excellent fit for QCLS among our scenarios of interest. It is flexible enough to be employed for ANSI as ambient noise data to be handled per node is not so large.

\section{Cross-Layer (PHY-MAC) Design}
\label{sec:cross}
A multi-objective network typically has to handle a complex data scheduling routine. In order to make our network design general purpose and flexible, we consider a network which is supposed to handle two data transmission tasks simultaneously. More specifically, data corresponding to an on-demand phenomenon (intermittent data) as well as a continuous data stream to be transmitted regularly. Later in Section~\ref{sec:appGroningen} we describe and analyze a practical scenario in Groningen in which such a general data scheduling becomes handy. These two different data streams for each sensor seem to call for two different wireless technologies/transmission modes. One approach to get around this issue is to handle both data streams within one data frame (equivalent of data packets over IP networks) \cite{TanenbaumCompNet} handled in a duty-cycled fashion. Note that the following approach is feasible when some delay in delivery of data streams is tolerable. If a considerable amount of on-demand data needs to be transmitted immediately, for instance in some early-warning systems (and in contrast to what we are interested in GMM), the following approach is not applicable. 

\begin{figure}[!t]
\centering
\includegraphics[width=0.8\textwidth]{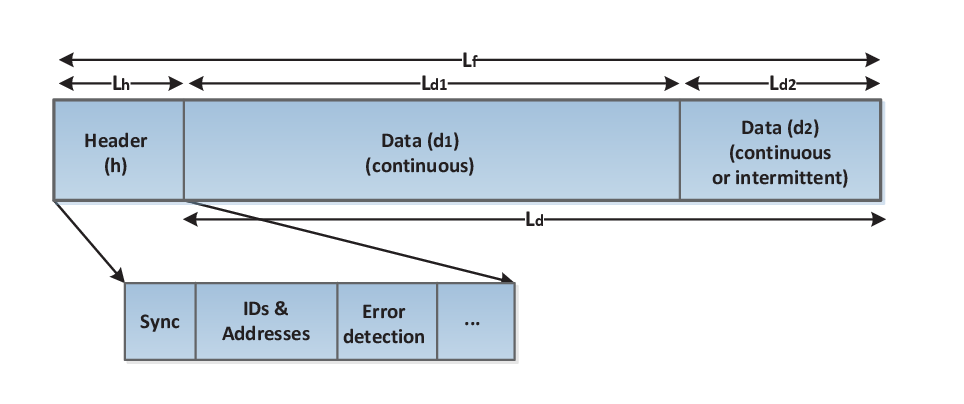}
\caption{Schematic frame format. Frame preamble is divided into two parts $d_1$ and $d_2$ where $d_2$ is filled in with on-demand data based on a trigger.}
\label{fig:FrameData}
\end{figure}

The goal in this section is not to delve into detailed MAC layer design of our data frames but to sketch a simplified view of the frame structure and discuss about its effect on our PHY layer design, i.e., the so-called cross-layer design considerations. Fig.~\ref{fig:FrameData} illustrates our approach where data frame is divided into two parts, header and main data. Frame header here basically refers to all the parts of an actual frame which contain no data but deal with other issues such as networking IDs and addresses, synchronization flags, and possible error detection/correction data. In order to accommodate both data streams, our data will contain two parts ($d_1$ and $d_2$) that will be filled in based on a trigger flag. As we briefly explained in Section~\ref{sec:phy}, consensus methods could be employed to avoid false triggers. If the trigger flag is ``$1$'', $d_1$ contains the continuous stream and $d_2$ contains a portion of the intermittent data. Bear in mind that we might need several frames in order to fully transmit the intermittent data corresponding to a trigger. On the other hand, if the trigger flag is ``$0$'' both $d_1$ and $d_2$ will contain the continuous data.  

Another important consideration in frame design is the proportion between the bytes allocated to header and data (preamble and payload). This is typically quantified as frame/protocol efficiency 
\begin{equation}
\label{etaf}
\eta_f = \frac{\text{data (payload) size}}{\text{total frame size}} = \frac{L_{d_1} + L_{d_2}}{L_f} = \frac{L_d}{L_f},
\end{equation}
where all the length values $L_{(.)}$ are in bytes. By looking at Fig.~\ref{fig:FrameData}, we have
\begin{equation}
L_f = L_{d_1} + L_{d_2} + L_h= L_d + L_h. 
\end{equation}
Typical values for $\eta_f$ in commercial networking standards stay above $0.8$ to ensure that a considerable amount of bandwidth is allocated to transmission of actual data. This is sometimes referred to as bandwidth-efficient transmission. The reason why we do not leave $d_2$ empty when trigger flag is ``0'' is to preserve frame efficiency (and thus, bandwidth-efficiency).
\begin{figure}[!t]
\centering
\includegraphics[width=0.95\textwidth]{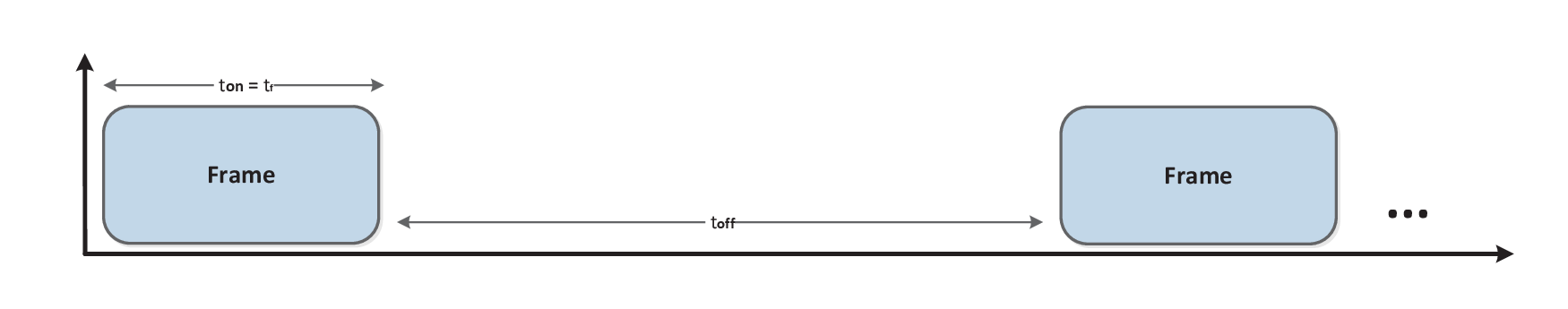}
\caption{Duty-cycled transmission protocol. Data will be sent in a frame and the channel will be free for a while after each transmission. }
\label{fig:dutyTrans}
\end{figure}

Power-saving strategies at the MAC level also impose other parameters which could be optimized. For different reasons including power efficiency in delay-tolerant applications as well as obeying certain channel occupancy restrictions for IoT-based networking, transceiver modules or M2M devices do not transmit data continuously but based on a duty-cycle. This is shown in Fig.~\ref{fig:dutyTrans} where in each cycle of duration $t_o = t_\text{on} + t_\text{off}$, the devices (our sensors) will transmit a frame of certain duration and then they go to the stand-by mode with transmission modules turned off leading to a considerably lower power consumption. Sometimes $t_\text{on} \ll t_\text{off}$, especially when a very long battery life-time is a major concern or hard channel occupancy limits are imposed by the governments in certain frequency bands (such as a $1$\% duty cycle for LoRa in Europe). 

We define duty-cycle as 
\begin{equation}
\label{dc}
\delta_c = \frac{t_f}{t_o} = \frac{t_f}{(t_f + t_\text{off})},
\end{equation}
where $t_\text{on} = t_f$. Based on the earlier explanations, we would like to formulate the relation between the required delay (denoted by $t_{D_2}$) to transmit the full data corresponding to each trigger (denoted by $D_2$), the transmission bit-rate (denoted by $R_b$), and data-splitting ratio defined as 
\begin{equation}
\begin{aligned}
\rho_d &= \frac{L_{d_1}}{L_{d_2}}\\ 
          &= \frac{L_{d_1} + L_{d_2} - L_{d_2}}{L_{d_2}} = \frac{L_{d}}{L_{d_2}} - 1. \label{rhod}
\end{aligned} 
\end{equation}
Let us compute the required time-delay for each $L_{(.)}$ as follows:
\begin{equation}
\label{ts}
\begin{aligned}
t_{d_1} = \frac{8 L_{d_1}}{R_b}, \, \, 
t_{d_2} = \frac{8 L_{d_2}}{R_b},\, \,
t_{f} = \frac{8 L_{f}}{R_b}, 
\end{aligned}
\end{equation}
all in seconds, where factor $8$ in the nominator accounts for the conversion of bytes to bits. Note that we need $\gamma$ full frames transmitted until the whole $D_2$ is delivered where
\begin{equation}
\label{gamma}
\begin{aligned}
\gamma &= \left \lceil \frac{L_{D_2}}{L_{d_2}} \right \rceil,\\ 
						 &= \left \lceil \frac{L_{D_2}}{\eta_f \, L_{f}} \, (\rho_d + 1) \right \rceil, 
\end{aligned}
\end{equation}
where $\lceil . \rceil$ stands for ceiling operator. In \eqref{gamma}, we have substituted $L_{d_2}$ from the \eqref{rhod} and then $L_d$ from \eqref{etaf} to arrive at the second line of \eqref{gamma}. 

Note that not all the frames arrive perfectly correct at the destination; especially in wireless channels always there is a percentage of frames $\lambda_f$ which are damaged. Typical values for $\lambda_f$ should be very small, such as $0.01$ to $0.05$. The damaged frames are detected using cyclic redundancy check (CRC) and depending on the MAC protocol typically a retransmission is requested for those frames. In highly-varying multipath wireless channels (for instance fast frequency-selective fading channels in harsh urban areas \cite{ProakisDigComm, GoldsmithWiComm}) typically more than one retransmission is required and that is why a precise estimate should take into account the channel statistics in order to compute the probability of frame damage and number of required re-transmissions. For the sake of simplicity of our estimations, we omit a probabilistic view of this problem and assume that a single re-transmission is enough to recover the damaged frames. Based on this assumption, we should send 
\begin{equation}
\label{gamma_bar}
\bar{\gamma} = \left \lceil \frac{L_{D_2}}{\eta_f \, L_{f}} \, (\rho_d + 1) \, (1 + \lambda_f) \right \rceil, 
\end{equation}
full frames to deliver whole $D_2$. Therefore, the total delay to transmit $D_2$ can be given by
\begin{equation}
\label{tD20}
\begin{aligned}
t_{D_2} &\approx \bar{\gamma} \, (t_f + t_\text{off}) = \bar{\gamma} \, (t_o),\\  
						&= \bar{\gamma} \, \left( \frac{8\, L_f }{R_b \, \delta_c} \right),
\end{aligned}
\end{equation}
where we have substituted $t_f$ from \eqref{ts} and $t_o$ from \eqref{dc}. Finally, by substituting \eqref{gamma_bar} into \eqref{tD20}, and solving it for $R_b$ we arrive at 
\begin{equation}
\label{Rb}
R_b = \frac{8 \, L_f}{\delta_c \, t_{D_2}} \, {\left \lceil \frac{L_{D_2} \, (1 + \rho_d) \, (1 + \lambda_f)}{\eta_f \, L_{f}} \right \rceil}.
\end{equation}
Looking at \eqref{Rb}, we notice that $R_b$ is inversely proportional to $t_{D_2}$, $\delta_c$ and $\eta_f$, and directly proportional to $\rho_d$ and $\lambda_f$. An appropriate $R_b$ should meet the following feasibility criteria:
\begin{itemize}
\item[-] It should provide us with a reasonably short $t_{D_2}$ to ensure that $D_2$ is already delivered before another trigger happens. So, $t_{D_2}$ and $R_b$ are entangled parameters which might require iterations to meet the requirements. 
\item[-] It should not be orders of magnitude smaller than data generation rate of the continuous data. This is to ensure that the sensors do not require a very large buffer size. 
\item[-] It should fit into typical data-rates being offered by available PHY technologies in the market, in our case LPWAN family. 
\end{itemize}
In order to find a proper value for $R_b$ we follow the design procedure sketched in Algorithm~\ref{alg:crossLayer}.
\begin{algorithm}[!t]
\caption{Bit-rate vs. total delay trade-off}\label{alg:crossLayer}
\begin{algorithmic}[1]
\State Estimate $L_{D_2}$, known a priori.  
\State Pick reasonable values for $\eta_f$, $L_f$, $\rho_d$, $\lambda_f$, and $\delta_c$.
\State Define a tolerable $t_{D_2}$. 
\While{$R_b$ does not fit into feasibility criteria}
\State Adjust $\rho_d$, $\eta_f$, $\delta_c$ (if possible), and re-iterate.
\State Compute $R_b$ from \eqref{Rb}.
\EndWhile
\end{algorithmic}
\end{algorithm}

\section{Design Study}
\label{sec:appGroningen}
The Groningen gas field in the north of the Netherlands is the largest gas producing field in Europe. Operated by the Nederlandse Aardolie Maatschappij (NAM), it has been in production since 1963. The production has resulted in subsidence and small-scale earthquakes. The surrounding area is mostly farm fields with limited or no wired/HDSL infrastructure. Besides, 4G coverage is relatively poor in major parts of the area, and more importantly as explained earlier, 4G is too power-consuming for minimum-maintenance long-range networks we envisage. This means Groningen could potentially be an opportunity for our IoT-based wireless networking ideas. Besides, comprehensive information regarding induced seismicity in Groningen can be found publicly available on NAM's website, as well as in several existing studies, for example in (\cite{van2015induced, de2015production}). The fact that such important input parameters are publicly available motivated us to conduct a design study on Groningen field. 

We investigate the feasibility of deploying a dense network of nodes based on our IoT-based wireless networking with two primary objectives: first, to conduct ANSI in order to obtain a better estimate of shallow subsurface velocities; second, to conduct GMM to obtain a better idea of the underlying source mechanism of the tremors. There are a few specifications concerning Groningen which highly affect our network design. In what follows, we revisit some of these specifications and explain how they play a role in our design. Next, we estimate data generation rates and volumes based on these pre-defined specifications. Finally, we pick appropriate network architecture(s) from our generic designs in Section~\ref{sec:phy}, and briefly look at the feasibility of employing them in Groningen.  

The area of interest is approximately $30$ km by $40$ km and this can even be further extended by considering a rim around the main gas field. On the south-western side of the covered area there is the city of Groningen, as well as there are a few villages located within the covered area which complicate our LoS communications. Most of the area are farmlands and there are no huge towers or other tall obstructions. Thus, our communications protocols for long-range transmissions in extreme situations should be able to mitigate non-LOS (NLoS) propagation due to blockage such as buildings and trees. The main wireless service providers in the region are T-Mobile, Vodafone, KPN, Telfort, and Tele2. The good news is that KPN is implementing LoRa networks throughout the Netherlands and it is not so expensive to ask for more gateways in Groningen to ensure a stable LoRa coverage. Interestingly, T-Mobile is also heavily investing on establishing a NB-IoT network inside and outside the Netherlands. NB-IoT offers about $20$ dB better coverage above 4G (which is poor in some regions as mentioned earlier), and thus ensures a much more stable connection throughout our whole area of interest in Groningen. 
\begin{figure}[!t]
 \centering
 \includegraphics[width=0.6\textwidth]{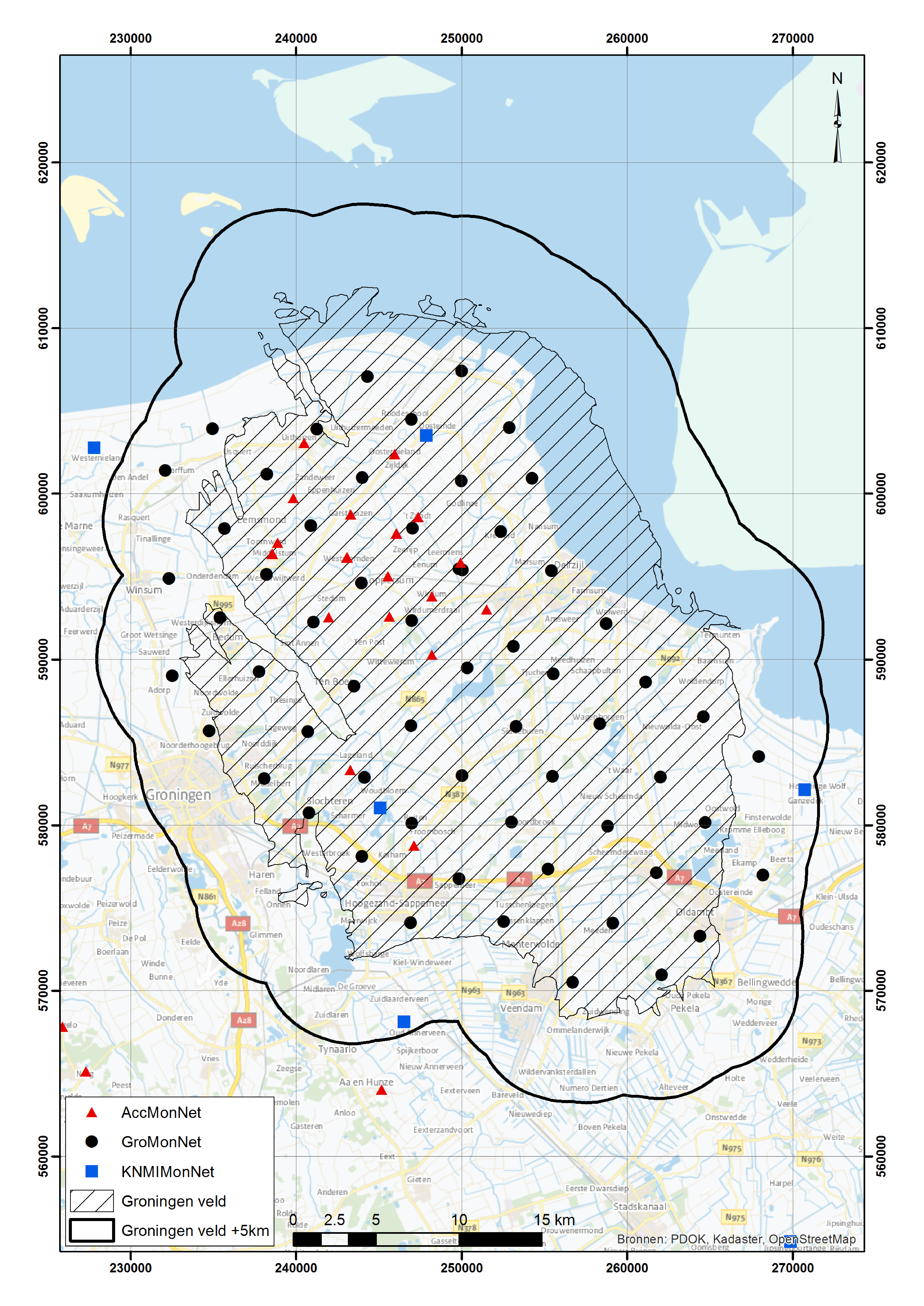}
\caption{Groningen field and KNMI borehole stations highlighted with black circles.}
\label{fig:Stations}
\end{figure}

Currently there exist a few other networks of sensors covering the region. One of them is the Koninklijk Nederlands Meteorologisch Instituut (KNMI) network which consists of about $80$ stations covering most of the province of Groningen. Each of these borehole stations (black circles in Fig.~\ref{fig:Stations}) has the same instrument configuration that is $4$ geophones in a $200$ meter deep well and an accelerometer at the surface. Most of the boreholes have access to HDSL (or Ethernet) high speed cable connections and unlimited power outlet, which we can make use of. We envisage a  network of sensors which is supposed to conduct a passive survey with two primary goals. First, ANSI to estimate shear-wave velocities in the subsurface based on ``continuous'' ambient noise measurements. Second, monitoring seismic activities in the subsurface showing up in the form of small-scale earthquakes which is a type of ``on-demand'' (triggered) operation. Especially for GMM, it is unnecessary that the recorded data at each sensor be transmitted to the FC for processing and decision making in a continuous and ``real-time'' fashion. This means that we can tolerate quite a bit of delay in this case. For the purpose of interferometry, we have to record/transmit continuous observations at each sensor. This again does not have to happen in a real-time fashion, and delays are tolerable as long as we transmit the data regularly. 

We make the following assumptions for our design. The sensors will be geophones, and are to be spread on an almost regular grid with an approximate spacing of $1$ km in each direction. Roughly speaking, for an area of about $40$ km by $40$ km, we need around $40 \times 40 = 1600$ sensors to be planted. This means that there is a good chance that some sensor locations fall within regions lacking cable Internet or with a poor 4G coverage. Therefore, a homogenous 4G network plan sounds to be unfeasible, and our network design should be flexible in this regard by thinking wider in terms of the technology to be employed. The candidate sensors are typically equipped with GPS that re-calibrates itself every few seconds to maintain accuracy. It is also recommended to upgrade the GPS to a network-wide differential GPS. Therefore, sensor locations are known up to about a meter accuracy, in the best case sub-meter accuracy. This known geographical location makes extracting meaningful data attached to physical locations from the network feasible. GPS can also provide us with time stamps with micro-second accuracy that can help us reach a time synchronization, if required. The network is expected to live with minimum maintenance and QC.  

\subsection{Data Generation Rates and Volumes}
\label{ssec:dataGen}
An important concern in our design is the data \emph{transmission} rate requirement for our network given sensor specifications and existing read outs. As we discussed in Section~\ref{sec:cross}, this in turn relates to the data \emph{generation} rates and acceptable latency to deliver data for analysis. We try to present rule of thumb computations in order to get an idea of the amount of data we are dealing with. Our computations are based on the following assumptions: 
\begin{itemize}
\item[I.] Sensors have $3$ components and a $24$ bit/sample ($3$-byte) precision.  
\item[II.] Sensors have an adjustable sampling rate between $50$ to $200$ sample/second.
\item[III.] For ANSI, we need to record continuously. To hamper the data generation rate, we only use the data associated with a single component of the sensors and just $4$ bits out of the $3$ bytes. This turns out to be accurate enough based on our investigations on the accuracy of estimated Green's functions. 
\item[IV.] For GMM, we intermittently record seismic events of considerable magnitude leading to an actual trigger. For each trigger, we have to record about $2$ minutes covering pre-event and post-event data.  
\end{itemize}

Based on these assumptions, we consider low, mid and high data generation rate scenarios for the two types of recordings, i.e., continuous for ANSI and intermittent for GMM:
\begin{subequations}
\begin{align}
4~\text{[bit/sample]} \times 1~\text{[component]} \times 50~\text{[samples/second]}  &= 200~\text{bps}, \label{LP1}\\
4~\text{[bit/sample]} \times 1~\text{[component]} \times 100~\text{[samples/second]}  &= 400~\text{bps},  \label{LP2}\\
4~\text{[bit/sample]} \times 1~\text{[component]} \times 200~\text{[samples/second} &= 800~\text{bps},  \label{LP3}
\end{align}
\end{subequations}
where \eqref{LP1} to \eqref{LP3} respectively correspond to low to high data generation rates for ANSI given different sensor sampling rates.  
\begin{subequations}
\begin{align}
3~\text{[byte/sample]} \times 3~\text{[component]} \times 100~\text{[samples/second]} &=  7.2~\text{kbps},  \label{HP1}\\
3~\text{[byte/sample]} \times 3~\text{[component]} \times 150~\text{[samples/second]} &=  10.8~\text{kbps},  \label{HP2}\\
3 ~\text{[byte/sample]} \times 3~\text{[component]} \times 200~\text{[samples/second]} &= 14.4~\text{kbps},  \label{HP3}
\end{align}
\end{subequations}
where \eqref{HP1} to \eqref{HP3} respectively correspond to low to high data generation rates for GMM given different sensor sampling rates. Now that we have two types of data corresponding to interferometry and monitoring to be delivered, we consider also two data streams for each sensor explained in the following. 

Continuous data stream for ANSI is to be transmitted at regular intervals as outlined for duty-cycled transmissions in Section~\ref{sec:cross} and shown in Fig.~\ref{fig:dutyTrans}. Considering \eqref{LP1} to \eqref{LP3}, and given the fact that we have to record continuously, we generate up to
\begin{subequations}
\begin{align}
365~\text{[days/year]} \times 24~\text{[hours]} \times 3600~\text{[second]} \times 200~\text{[bps]} &\approx 788.4~\text{MB}, \label{LPS1}\\ 
365~\text{[days/year]} \times 24~\text{[hours]} \times 3600~\text{[second]} \times 400~\text{[bps]} &\approx 1.577~\text{GB}, \label{LPS2}\\
365~\text{[days/year]} \times 24~\text{[hours]} \times 3600~\text{[second]} \times 800~\text{[bps]} &\approx 3.154~\text{GB}, \label{LPS3}
\end{align}
\end{subequations}
of data per sensor over a year time. In order to differentiate between bit and byte we use capital B for the latter. 
\begin{figure}[!t]
 \centering
 \includegraphics[width=0.9\textwidth]{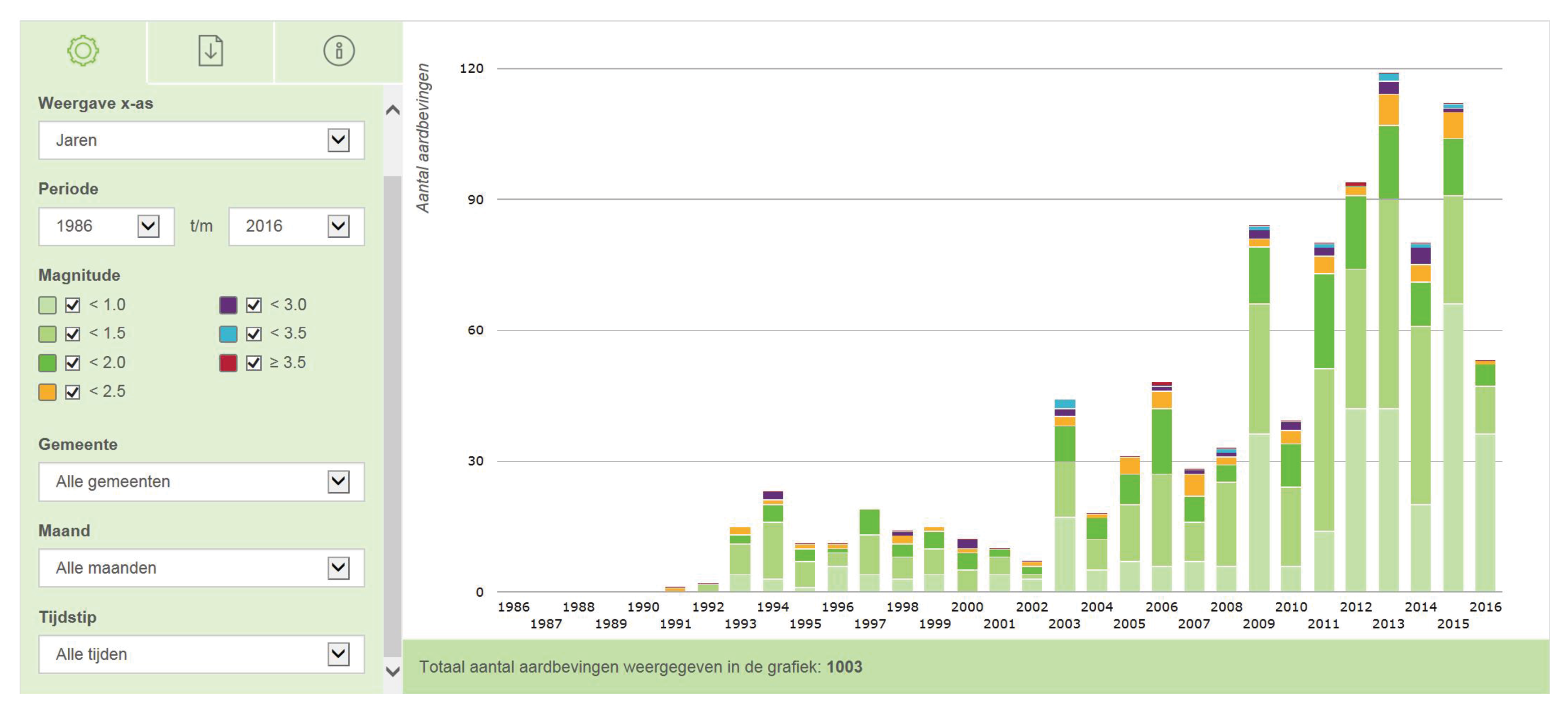}
\caption{Earthquakes in Groningen field from 1986 to 2016, a snapshot taken from NAM's observation portal.}
\label{fig:NAMportal}
\end{figure}

Delay-tolerant data stream for GMM is to be transmitted at intermittent intervals in an on-demand fashion based on triggers. Let us consider \eqref{HP1} to \eqref{HP3}, and 2 minutes of recording per trigger. Based on the statistics from KNMI illustrated on NAM's observation portal shown in Fig.~\ref{fig:NAMportal}, during the past three decades the maximum number of earthquakes (about $120$) within the range of $1$ to $3.5$ Richter have been detected in 2013. We thus consider a worst case scenario of $500$ triggers per year per sensor for the whole network. Note that is practice all these worst case $500$ tremors might not necessarily be detected and thus trigger all the sensors but a group of them depending on the regional proximity to the sensors. Taking this into account, will ensure that $500$ triggers is a solid worst case scenario. This leads to  
\begin{subequations}
\begin{align}
500~\text{[triggers]} \times 120~\text{[second]} \times 7.2~\text{[kbps]} &= 54~\text{MB}, \label{HPS1}\\  
500~\text{[triggers]} \times 120~\text{[second]} \times 10.8~\text{[kbps]} &= 81~\text{MB}, \label{HPS2}\\ 
500~\text{[triggers]} \times 120~\text{[second]} \times 14.4~\text{[kbps]} &= 108~\text{MB}, \label{HPS3} 
\end{align}
\end{subequations}
of data per sensor over a year time. This is based on a worst-case assumption that a trigger event will result in the entire network reading out. It is possible to avoid reading out the entire network for smaller events as well as for significant number of expected false triggers through more intelligent readout protocols. For a network of $1600$ sensors, taking into account the mid cases \eqref{LPS2} and \eqref{HPS2} among the previous estimations for both streams, we generate approximately
\begin{equation}
(81 + 1577)~\text{[yearly data per sensor]} \times 1600~\text{[sensors]} \approx 2.66 ~\text{TB},
\end{equation}
of data for the whole network over a year time. TB here stands for Tera bytes. Note that we do not deal with a ``Big Data'', which is a good news for our low-data-rate IoT-based wireless technologies. 

\subsection{Proposed Network Architecture}
\label{ssec:propArch}
%
\begin{table}[!t]
\renewcommand{\arraystretch}{1.2}
\caption{Cross-Layer Design Parameters}
\label{tab:params}
\centering
{\small
\begin{tabular}{| m{1.8cm} |c | c | c |}
\hline
\textbf{Parameters} & Low & Mid &  High \\
\hline
\hline 
$\eta_f$ & 0.9 & 0.95 & 0.98 \\
\hline 
$L_{D_2} $ & $108$ kB  &  $162$ kB  &  $216$ kB \\
\hline
$L_f $ & $64$ B  &  $128$ B  &  $256$ B\\
\hline
$t_{D_2}$ & $1$ hour  &  $10$ hours  &  $1$ day \\
\hline
$\delta_c$ & 0.01  &  0.05  & 0.1 \\
\hline
$\rho_d$ & $1$  &  $3$  &  $5$  \\
\hline
$\lambda_f$ & $0.01$  &  $0.05$  &  $0.1$  \\
\hline
\end{tabular}}
\end{table}
Now that we have an estimate of size and rate of the ``generated'' data per sensor and across the whole network, we can use our simplified model in Section~\ref{sec:cross} to get an estimate of the required ``transmission'' data-rate. We have collected the estimated data in the previous subsection along with some other practical values in Table~\ref{tab:params}. We follow Algorithm~\ref{alg:crossLayer} in order to compute an appropriate $R_b$ meeting our feasibility criteria. Let us pick some reasonable values from the table and set $\eta_f = 0.9$, $L_{D_2} = 216$ kB, $L_f = 128$ B, and $\lambda_f = 0.01$. We decide to dedicate $1/2$ of the payload to the high-precision data, i.e., $\rho_d = 1$, and consider a tolerable delay of $t_{D_2} = 10$ hours in order to make sure strong ground motions are completely reported in less than a day. We also set $\delta_c = 0.01$ which can even pass the extreme duty-cycle restrictions that LoRaWAN has to meet in Europe. All in all, \eqref{Rb} yields
\begin{eqnarray}
R_b &=& \frac{8 \, L_f}{\delta_c \, t_{D_2}} \, {\left \lceil \frac{L_{D_2} \, (1 + \rho_d) \, (1 + \lambda_f)}{\eta_f \, L_{f}} \right \rceil} \nonumber\\
       &=& \frac{8 \times 128 }{0.01 \times 10 \times 3600} \, {\left \lceil \frac{216000 \times (1 + 1) \, (1 + 0.01)}{0.9 \times 128 } \right \rceil} = 10.77 \, \text{kbps}.
\end{eqnarray}
This is reasonably low to be realized with an affordable low-bandwidth technology in the market such as LoRa, NB-IoT, etc., it is higher than data generation rate of ANSI not to impose any further delay in delivering continuous stream of data, and also delivers GMM data within $10$-hour time which is fine from an operational perspective. 

Suppose we need to deliver GMM high precision data with less delay, say $1$ hour. In such a case, we need to keep $\rho_d = 1$ as increasing it would have inverse effect. As a result, we need $10$ times larger data-rate 
\begin{eqnarray}
R_b &=& \frac{8 \times 128 }{0.01 \times 1 \times 3600} \, {\left \lceil \frac{216000 \times (1 + 1) \, (1 + 0.01)}{0.9 \times 128 } \right \rceil} = 107.7 \, \text{kbps},
\end{eqnarray}
making it unfeasible to be realized with LoRa but still doable with NB-IoT even with $\delta_c = 0.01$. A possible solution to make it still work for LoRa is to use $\rho_d < 1$ and increase $\delta_c$ using different LoRa sub-bands to tackle duty-cycle restrictions. LoRa can transmit in three main sub-bands and a fourth separate band which in total can potentially sum up to $12.1\%$ duty-cycle. Nonetheless, we have a few tuning parameters that we can adjust to suit our feasibility criteria. 

Next steps are to pick a proper PHY wireless technology and lay out a schematic network architecture. Now that $R_b$ is not so restrictive and the Groningen area owns a reasonable wireless infrastructure, we pick two technologies from the two categories of LPWAN, i.e., LoRa and NB-IoT. As we explained earlier, there is growing potential for both technologies in the Netherlands which further supports our choices of technologies. As a result of our choices, we can propose two network architectures based on both hybrid private-cellular and pure cellular designs explained in Section~\ref{sec:phy}. More specifically, we propose to modify Architectures I and II to fit in our particular scenario of interest as follows. 

Given the growth of NB-IoT in major telecommunications play-makers' plan in the Netherlands and its promising $20$ dB extended coverage above 3G/4G, Architecture I can be put in action with minimum modifications. This architecture equipped with NB-IoT technology can easily handle the large number of nodes as well as the relatively large area of the Groningen field. The nodes should obviously be equipped with NB-IoT wireless transceiver modules and the boreholes with HDSL connection will play the role of wired nodes. The latter can at the same time act as data relays, if necessary. A combination of the low data-rate requirement per sensor, high capacity of the LoRa gateways, and availability of cellular backhaul in the region, suggests that a modified version of Architecture II should also be feasible for Groningen. Notably, the HDSL connections available at the well-sites in Groningen makes them an excellent choice to place the LoRa gateways, and thus directly forward the data through the IP network. Given the distribution of the well-sites in the area, the LoRa-equipped nodes only need to be able to connect to a gateway located within a few kilometers. Based on experiments we have conducted with LoRa transceivers (briefly explained in the next section), even in urban areas good signal reception is guaranteed within few kilometers. Despite that specific locations might exist where additional LoRa or NB-IoT gateway has to be placed. In such a case, there are gateways in the market that allow for direct connection to the closest 3G (or 4G) wireless tower or Node-B in the vicinity and from there to the backhaul network. 

\subsection{Considerations for Total Cost of Operation}
\label{ssec:costNet}
%
\begin{table}[!t]
\renewcommand{\arraystretch}{1.2}
\caption{Cost Estimate Summary}
\label{tab:costSum}
\centering
{\footnotesize
\begin{tabular}{| m{4cm}| *{2}{c|}*{2}{c|}}
\hline
\textbf{Cost Co-Factors} & \multicolumn{2}{c|}{LoRa Network} &  \multicolumn{2}{c|}{NB-IoT Network} \\
\hline
\hline 
End node [no, price]  & $1600$ & $10 \, \$$ & $1600$ & $5 \, \$$ \\
\hline 
Gateways [no, price] & $45$ & $1,000 \, \$$ & $-$ & $-$ \\
\hline
Extra mast [no, price] &  $-$ & $-$ & $5$ & $20,000 \, \$$ \\
\hline
Subscription [node/year] & \multicolumn{2}{c|}{$-$} & \multicolumn{2}{c|}{$30 \, \$$}\\
\hline
Total Opex [year] & \multicolumn{2}{c|}{$61,000 \, \$ $} & \multicolumn{2}{c|}{$156,000 \, \$ $} \\
\hline
\end{tabular}}
\end{table}
Establishing a network and its maintenance involves different costs. We briefly look into a few operational expenditure (Opex)-related factors and their rough estimates for LoRa and NB-IoT technologies. Typically, exact pricing values are unclear until the final agreement with the implementing company is reached, especially for evolving technologies such as NB-IoT. Therefore, we emphasize that the following data is just to provide rough estimates. Also, note that several maintenance-related costs such as personnel and vehicles in the face of day-to-day operational issues (such as equipment failure, land owner problems, and vandalism) are omitted here, as they apply to both network designs. 

As mentioned earlier, we consider approximately $1600$ nodes spread over a region of about $40$ km by $40$ km. As explained in the previous subsection, we consider LoRa gateways to be placed at the well-sites which are approximately space by $6$ kilometers. So, we roughly need $\lceil 40/6 \times 40/6 \rceil = 45$ LoRa gateways. NB-IoT relies on existing cellular infrastructure, that is why we considered a worst-case scenario of requiring $5$ extra wireless masts to be put up, which can possibly even be dropped. The subscription is required for NB-IoT nodes, but also could be required for large-scale LoRa networks tapping into the existing infrastructure for instance provided by KPN in Groningen. In such a case the gateway costs associated with LoRa should be replaced with yearly subscription rates. Putting all the numbers together as summarized in Table~\ref{tab:costSum}, the total Opex of establishing the NB-IoT turns out to be more than twice the LoRa network for a one-year run. If NB-IoT backbone network is mature enough in a region and extra masts are no longer necessary, this relative difference in total cost would drop drastically. It is also worthy of being emphasized that the cellular-based nature of NB-IoT (as compared to a private LoRa network considered here) has the potential to provide a higher data-rate and a more stable performance in terms of data delivery.  

\section{Seismic QC Field Test with LoRa-Enabled Nodes}
\label{sec:loraPoC}
%
\begin{figure}[!t]
 \centering
 \includegraphics[width=0.9\textwidth]{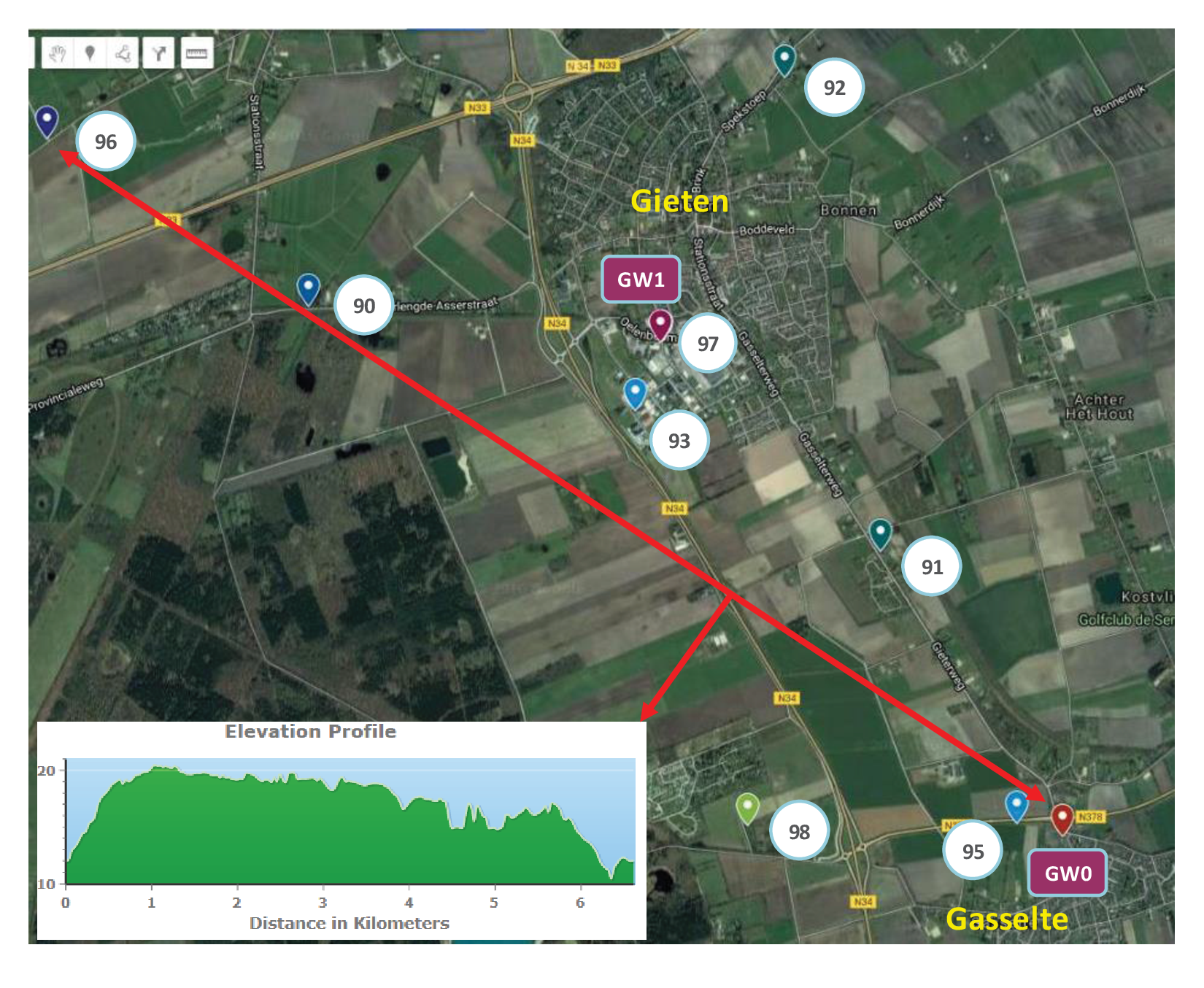}
\vspace{-0.3cm}
\caption{Area of the PoC field test in Drenthe, topography of the region, nodes ($90$ to $98$ with the exception of $94$ which is our mobile node) and gateways (GW0 and GW1) locations. To give the reader an idea of the scales, GW0 and node $96$ are placed $6.5$ km apart, and the elevation profile on the bottom left shows a NLoS condition and considerable elevation variations between the node and the gateway.}
\label{fig:LoRa_Trial_2}
\end{figure}
%


As a predominant member of LPWAN family, LoRa offers a very compelling mix of long range, low power consumption and secure data transmission. It is easy to plug into the existing infrastructure and offers a solution to serve battery-operated IoT applications \cite{LoRa}. LoRaWAN is a protocol specification built on top of the LoRa technology developed by the LoRa Alliance. It uses unlicensed radio spectrum in the ISM bands to enable wide area communication between remote sensors and gateways connected to the backbone network \cite{LoRaWAN}. 

As the first step to materialize our vision on real-time IoT-based wireless seismic, in December 2016, Shell in collaboration with third party Innoseis ran a proof-of-concept (PoC) field test in the province of Drenthe in the north of The Netherlands. To this aim, LoRa-enabled nodes with embedded RN2483 LoRaWAN chips with optimized sensing technology and packaging were produced. The main goal of this test was to assess the fundamental performance of a LoRa-based wireless seismic network for real-time seismic QC/monitoring. The field test had three main stages: first, performance assessment of a single transceiver pair of a LoRa-enabled seismic node and a LoRa gateway (Kerlink Wirnet Station), second, network level performance with $8$ nodes and $2$ gateways, and third, another network-level test where a mobile nodes was added to the network. 

Fig.~\ref{fig:LoRa_Trial_2} shows the coverage area of about $13~\text{km}^2$ as well as distribution of the nodes and gateways. The first gateways (GW0) was installed on top of a drilling facility at a height of $11$ m in Gasselte, and the second one (GW1) was mounted on a pole at a height of $4.5$ m in Gieten. In the figure, the nodes are numbered from $90$ to $98$ where $94$ is assigned to the mobile node which is not shown in Fig.~\ref{fig:LoRa_Trial_2} and is only introduced in the third stage. It is worth highlighting that the coverage area involved a lot of NLoS complications including trees and buildings, and the weather was partly rainy and foggy, and at times freezing cold during the test. The network was deployed in less than one day, and the whole test was run for three days after which the network was easily retrieved. 

\begin{figure}[!t]
 \centering
 \includegraphics[width=0.9\textwidth]{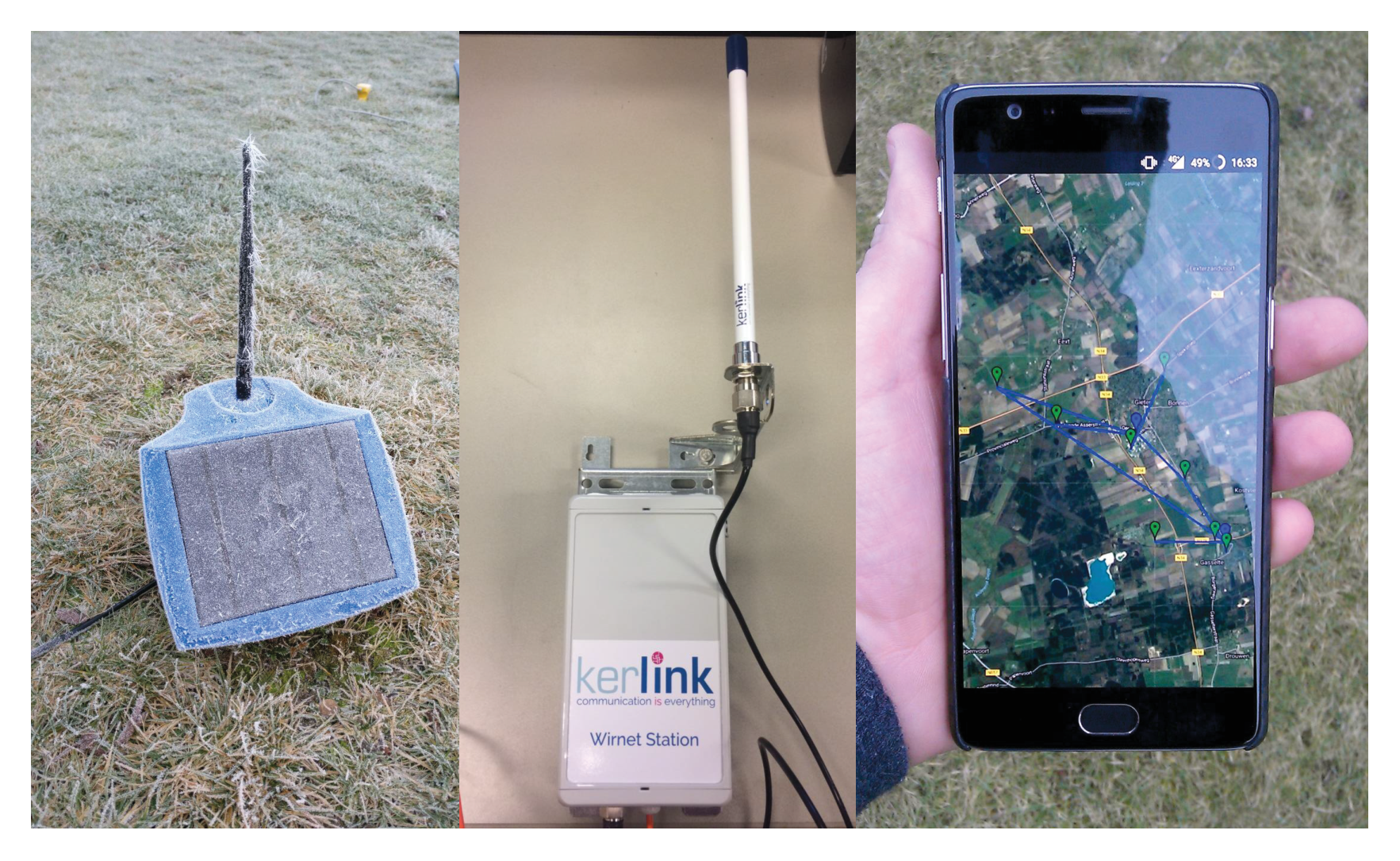}
\caption{On the left, LoRa-enabled with solar panel on top in freezing temperature of December; in the middle, Kerlink Wirnet Station providing direct 3G/4G connection as well as Ethernet connection possibilities, GPS with sub-meter location and microsecond time accuracies, and power on Ethernet (PoE) power supply; on the right, application tool running on a cell-phone providing real-time visual QC and status information.}
\label{fig:TN_Kerlink_App}
\end{figure}

We have employed the public SEMTECH network server for the field test allowing us to keep track of also and download detailed information about communicated packets in a real-time fashion on the cloud. The network server, which was accessible via Internet on authorized personal devices, also acted as our data server where we could download and analyze the data itself.  Fig.~\ref{fig:TN_Kerlink_App} show the our LoRa-enabled solar-paneled node, the employed gateway, and our custom-developed application providing us with visual QC tools for real-time monitoring.

We have conducted extensive tests in three stages in order to explore the potentials as well as shortcomings of such a LoRa-based wireless seismic network. In the first stage our focus was on single node-gateway performances between all possible node-gateway pairs. For instance, for a $3.2$ km distance between GW0 and node $93$ in an LoS condition an average frame error rate (FER), computed using CRC, as low as $5.2\%$ was achieved in spite of about 40\% of the first Fresnel zone being obstructed by the elevation profile. In a more assertive attempt, for a $6.5$ km partly obstructed NLoS scenario, FER has increased to about $40\%$ which is still good given that there was only a single gateway (a single NLoS communication link). 

Our measurements also show that the communication modules of the LoRa-enabled node obeying the $1$\% duty-cycle restriction of LoRa in Europe used up on average less than a milliWatt power which would allow our standard $48$ Watt-hour (Wh) batteries (excluding possible recharge through solar panels) to survive more than 6 years. This corroborates the fact that LoRa can be employed for long-lasting (seismic) operations without imposing a major power-consumption load, as opposed to many wireless seismic technologies in the market for which the wireless modules are power consuming. In the second stage, where proper LoRaWAN network has been put in place, adding only one more gateway has resulted in a FER less than $10\%$ by all the nodes seen by both gateways, which clearly highlights the importance of a proper network architecture design. We have also shown that our nodes could nicely mitigate interference from other LoRa users, as well as interference sources working around the same frequency band ($868$ MHz). The third stage of the PoC field test was partly devoted to a mobility and handover test with node $94$ mounted on a driving car. We have illustrated that LoRa data frames could be seamlessly received and decoded by the gateways even from our mobile node ($94$) moving as fast as $100$ km/h, which by itself highlights the potential of LoRa for a wider range of applications within the Oil and Gas industry including asset-tracking. 

In a nutshell, our field test results corroborate that cheap (less than 10 USD) subscription-free LoRa chips can be embedded into our seismic sensory systems allowing us to transmit more that $6$ MB of data per node per day over distances of a few kilometers while the data could be monitored real-time on cloud. Note that $6$ MB of data can only be handled if all the frequency sub-bands of LoRa (adding up to $12.1\%$ allowed duty-cycle) are employed at the highest data-rate offered by its lowest spreading factor; employing only a single frequency sub-band will result in a fraction of this amount. Nonetheless, it also means that a properly-engineered network of $1600$ nodes can potentially handle $6 \times 365 \times 1600 \approx 3.5$ TB of data which is more than what we have estimated for our network design in Groningen (see Subsection~\ref{ssec:dataGen}), and thus re-confirms that if our applications are delay tolerant such an IoT-based network can efficiently handle the data aggregation and transmission. Interested reader is referred to our extended work focusing on this PoC field test in \cite{LeadingEdge17}.

\section{Concluding Remarks}
\label{sec:conc}
%
Wireless seismic technologies are being picked up by the market with an unprecedented rate in the past few years. This is because compared to the traditional cable seismic they offer a more cost-efficient solution with less environmental impact without the complications of transporting, maintaining and retrieving cable-based systems; they are less demanding in terms of maintenance and provide the possibility of real-time data acquisition. 

We have observed the advent of a new generation of wireless technologies (so-called LPWANs) with inherent Internet of things (IoT) compatibilities and we have set them at the core of our networking design. Our proposed design combines affordable low-power long-range wireless technologies, advanced and scalable networking protocols, and Internet of sensors with cloud computing for storage and processing. The result is a plug and play network where anyone can define/add new sensors; it can operate in a (near) real-time fashion to address a wide variety of demands, and it will be scalable to thousands of sensors with worldwide accessibility to the acquired data. We have proposed two IoT-based wireless networking architectures based on different categories of LPWANs and have matched them with our seismic scenarios of interest. We then have presented a practical study on Groningen field where we have incorporated our design architectures and have corroborated our detailed networking quantitative estimates. Finally, we have put an step further in materializing our vision of IoT-based wireless seismic by conducting a proof-of-concept field test with LoRa and have presented promising first results for seismic QC. 

Even though the main scenarios discussed in this paper are about seismic applications, we believe the underlying concept revolving around a flexible IoT-based wireless network of cheap nodes making use of cloud services is applicable to a wide variety of applications in the Oil and Gas industry, from upstream (seismic interferometry) to mid-stream (pipeline monitoring) and even downstream (asset tracking). Our future effort will be focused on large-scale implementation of IoT-based wireless technologies (LoRa and NB-IoT) for such scenarios. 

\newpage
\section{Acknowledgments}
The authors would like to thank Shell Global Solutions International B.V. for permission to publish this paper. We would like also to thank Ian McKay, Wim Walk, Jeff Jackson, Dirk Smit (Shell Global Solutions International B.V.) for their guidance and support, and Jo van den Brand and Mark Beker (Innoseis B.V.) for their collaborations.    

\clearpage
\small
\bibliography{fullRef}

\bibliographystyle{gp}  

\end{document}